\begin{filecontents}{}
gsave
72 31 moveto
72 342 lineto
601 342 lineto
601 31 lineto
72 31 lineto
showpage
grestore
\documentclass[epj]{svjour}
\usepackage{array}
\usepackage{amsmath}
\usepackage{amssymb}
\usepackage{graphics}
\usepackage{graphicx}
\usepackage{mathrsfs}

\def\rhoz{\rho_{\raisebox{-0.75pt}{\tiny 0}}}
\def\barrhoz{\bar{\rho}_{\raisebox{-0.75pt}{\tiny 0}}}
\def\epsz{\varepsilon_{\raisebox{-0.75pt}{\tiny 0}}}
\def\alphad{\alpha_{\raisebox{-1pt}{\tiny  D}}}
\def\alphadDM{\alpha_{\raisebox{-1pt}{\tiny  D}}^{\raisebox{0pt}{\tiny DM}}}
\def\rskin{r_{\rm skin}}

\begin{document}

\title{Symmetry Energy Constraints from Giant Resonances:\hfill\\A Theoretical Overview}
\author{J. Piekarewicz}
\institute{Department of Physics \\
              Florida State University \\
              Tallahassee, FL 32306-4350 \\
              USA}	
\date{Received: date / Revised version: date}
%
\abstract{
Giant resonances encapsulate the dynamic response of the nuclear ground state 
to external perturbations. As such, they offer a unique view of the nucleus that is 
often not accessible otherwise. Although interesting in their own right, giant 
resonances are also enormously valuable in providing stringent constraints on 
the equation of state of asymmetric matter. We this view in mind, we focus on 
two modes of excitation that are essential in reaching this goal: the isoscalar 
giant monopole resonance (GMR) and the isovector giant dipole resonance (GDR). 
GMR energies in heavy nuclei are sensitive to the symmetry energy because 
they probe the incompressibility of neutron-rich matter. Unfortunately, access 
to the symmetry energy is hindered by the relatively low neutron-proton asymmetry 
of stable nuclei. Thus, the measurement of GMR energies in exotic nuclei is 
strongly encouraged. In the case of the GDR, we find the electric dipole 
polarizability of paramount importance. Indeed, the electric dipole polarizability 
appears as one of two laboratory observables---with the neutron-skin 
thickness being the other---that are highly sensitive to the density dependence 
of the symmetry energy. Finally, we identify the softness of skin and the nature 
of the pygmy resonance as important unsolved problems in nuclear structure.
} 
\maketitle
\section{Introduction}
\label{intro}

Nuclear saturation, the existence of an equilibrium density, is a
hallmark of the nuclear dynamics. That the size of the nucleus 
increases as $A^{1/3}$ is one of the best-known consequences 
of nuclear saturation. The venerable semi-empirical mass
formula of Bethe and Weizs\"acker, conceived shortly after the
discovery of the neutron by Chadwick, treats the nucleus as an
incompressible quantum drop consisting of $Z$ protons and $N$ neutrons
($A\!=\!Z\!+\!N$). The mass formula may be written in general in terms
of the individual nucleon masses ($m_{p}$ and $m_{n}$) and the nuclear
binding energy $B(Z,N)$ that contains all the complicated nuclear
dynamics: $M(Z,N)\!=\!Zm_{p}\!+\!Nm_{n}\!-\!B(Z,N)$. In the context 
of the liquid-drop formula the binding energy is written in terms of a
handful of empirical parameters that portray the physics of a quantum 
drop.  That is,
\begin{equation}
 B(Z,N)\!=\!a_{{}_{{\rm V}}}A - a_{{}_{{\rm S}}}A^{2/3} - 
 a_{{}_{{\rm C}}}\!\frac{Z^{2}}{A^{1/3}} - a_{{}_{{\rm A}}}\!\frac{(N\!-\!Z)^{2}}{A}+\ldots
 \label{BWMF}
\end{equation}
The volume term represents the binding energy of a large and symmetric
($Z\!=\!N\!=\!A/2\!\gg\!1$) system in the absence of Coulomb
forces. The next three terms are correction terms due to the
development of a nuclear surface, the Coulomb repulsion among
protons, and the Pauli exclusion principle that favors symmetric
systems. Note that in both the surface and Coulomb terms one has 
already used nuclear saturation to write the radius of the nucleus in 
terms of the equilibrium (or saturation) density 
$\rhoz\!=\!0.148\,{\rm fm}^{-3}$ as follows:
\begin{equation}
 R(A)=r_{{}_{\!0}}A^{1/3}\,, \;{\rm where}\;
 r_{{}_{\!0}}\!=\!\sqrt[3]{\frac{3}{4\pi\rhoz}}\!=\!1.17\,{\rm fm}\,.
 \label{Radius}
\end{equation}
Although such a smooth formula gives a remarkably good description of
the masses of stable nuclei, it is unable to account for local
fluctuations associated with the emergence of nuclear shells and the
concomitant appearance of magic numbers. To overcome such a drawback,
a {\sl macroscopic-microscopic} approach was developed, where the
nuclear binding energy is separated into two components: one large and
smooth (as in the liquid drop model) and the other one small and
fluctuating to properly account for shell effects.  The
macroscopic-microscopic approach has enjoyed its greatest success in
the work of M\"oller and
collaborators\,\cite{Moller:1993ed,Moller:1997bz,Moller:2012}, and
Duflo and Zuker\,\cite{Duflo:1994,Zuker:1994,Duflo:1995}. Although
refinements to the mass formula have been made to meet new and
increasing challenges, the structure of this 75 year-old formula
remains practically unchanged.

We now consider the thermodynamic limit of the liquid-drop formula in
which both the number of nucleons and the volume are taken to
infinity, but their ratio remains fixed at the saturation
density. Moreover, we ignore the electroweak sector so that Coulomb
forces are absent and both $Z$ and $N$ are individually conserved. In
this limit the binding energy per nucleon may be written as
\begin{equation}
 {\cal E}(\alpha) \equiv -\frac{B(Z,N)}{A} = \epsz + J\alpha^{2} \;,
 \label{EAsym}
\end{equation}
where $\epsz\!=\!-a_{{}_{{\rm V}}}$, $J\!=\!a_{{}_{{\rm A}}}$, and 
$\alpha\!=\!(N-Z)/A$ is the neutron-proton asymmetry. This simple 
formula suggests that a large symmetric liquid drop of density
$\rhoz$ has a binding energy per nucleon of  $\epsz\!\approx\!-16$\,MeV 
and that there is an energy cost of $J\!\approx\!32$\,MeV in converting 
all protons into neutrons, namely, in turning symmetric nuclear matter into 
pure neutron matter. However, in reality the liquid drop is not incompressible, 
so the semi-empirical mass formula fails to capture the response of the
liquid drop to density fluctuations. Given that the liquid-drop formula 
provides a remarkably good description of the masses of stable nuclei, 
it is evident that nuclear masses alone, while highly sensitive to $\epsz$ 
and $J$, are insensitive to the density dependence of these parameters. 
Note that the equation of state (EOS) $E(\rho,\alpha)$ dictates how the energy
per nucleon changes as a function of both the density and the neutron-proton 
asymmetry. In essence then, the masses of stable nuclei provide meaningful 
constrains only for $E(\rho\!\approx\!\rhoz,\alpha)$. 

To probe the density dependence of the EOS one must study the response
of the liquid drop to external perturbations. That is, one uses an external probe 
to drive the system away from equilibrium and then records how the system 
responds to such a perturbation. In this contribution we will focus on two 
particular nuclear excitation modes: the isoscalar monopole resonance and 
the isovector dipole resonance\,\cite{Harakeh:2001}. For the impact of other 
excitation modes in constraining the nuclear equation of state see  
Refs.\,\cite{Paar:2007bk,Sagawa:2007pi,Roca-Maza:2013yha}, 
references contained therein, and the contribution to this volume by Col\`o, 
Garg, and Sagawa. 

The isoscalar monopole resonance measures the collective response of the 
nucleus to density fluctuations. Pictorially, this collective excitation in which 
protons and neutrons oscillate in phase around the equilibrium density may be 
perceived as a spherical {\sl breathing} mode. Given that nuclear matter 
saturates, the pressure at saturation density vanishes. Thus, the isoscalar 
monopole resonance probes the curvature of the EOS at saturation density, 
or equivalently, the incompressibility of neutron-rich matter\,\cite{Piekarewicz:2008nh}. 
Naturally, the incompressibility of asymmetric matter is sensitive to the symmetry 
energy. However, a more direct probe of the density dependence of the symmetry 
energy is the isovector dipole resonance. One may visualize this mode of excitation 
as an out of phase displacement of neutrons relative to protons. Given that this 
oscillation results in the formation of two dilute quantum fluids---one neutron rich 
and the other one proton rich---the symmetry energy acts as the restoring force. 
Note that in contrast to the EOS of symmetric nuclear matter, the symmetry pressure 
does not vanish at saturation density. An enormous effort has been devoted 
in recent years to determine the symmetry pressure.

Even though our contribution to this special volume is limited to giant
resonances, we find useful to discuss briefly the impact of other
observables in constraining the nuclear symmetry energy. Although the
symmetry pressure (or equivalently the slope of the symmetry energy
at saturation density $L$) is not an observable, it has been shown
to be strongly correlated to the neutron-skin thickness of
${}^{208}$Pb\,\cite{Brown:2000,Furnstahl:2001un,Centelles:2008vu,RocaMaza:2011pm}.
The neutron-skin thickness is defined as the difference between 
the neutron ($r_{n}$) and proton ($r_{p}$) root-mean-square radii.
The charge (and proton) distribution in nuclei is known with exquisite 
accuracy due to the pioneering work of Hofstadter in the late 
1950's\,\cite{Hofstadter:1956qs} and that continues to this day 
with the advent of powerful continuous electron beam facilities.  
Instead, challenging  parity-violating experiments are required to 
cleanly measure neutron densities \,\cite{Donnelly:1989qs}.  
Indeed, because the weak charge of the neutron is much larger than 
that of the proton, parity-violating electron scattering provides a clean 
probe of the neutron radius that is free from large and uncontrolled
strong-interaction uncertainties. As one combines these results with
the highly-precise measurements of nuclear charge
radii\,\cite{Angeli:2013}, one obtains a fairly model-independent
determination of the neutron-skin thickness of ${}^{208}$Pb, and thus
a stringent constraint on the density dependence of the symmetry
energy. We note that the Lead Radius Experiment (``PREX'') at the
Jefferson Laboratory has provided the first model-independent evidence
on the existence of a neutron-rich skin in 
${}^{208}$Pb\,\cite{Abrahamyan:2012gp,Horowitz:2012tj}. Through the
use of some mild assumptions PREX determined the neutron-skin of 
${}^{208}$Pb to be\,\cite{Abrahamyan:2012gp}:
\begin{equation}
 r_{\rm skin}^{208}\!\equiv\!r_{n}^{208}\!-\!r_{p}^{208}\!=\!{0.33}^{+0.16}_{-0.18}\,{\rm fm}.
\label{rskin208}
\end{equation}
For a detailed account of this significant achievement and efforts to
measure the neutron-skin thickness of ${}^{48}$Ca, see the contribution 
to this volume by Horowitz, Kumar, and Michaels.
An accurate determination of the neutron radius of 
${}^{208}$Pb---and the resulting constrain on $L$---represents a critical 
milestone with far reaching implications in areas as diverse as nuclear
structure~\cite{Brown:2000,Furnstahl:2001un,Centelles:2008vu,RocaMaza:2011pm},
atomic parity
violation~\cite{Pollock:1992mv,Sil:2005tg,Guena:2005uj,Behr:2008at},
heavy-ion
collisions~\cite{Tsang:2004zz,Chen:2004si,Steiner:2005rd,Shetty:2007zg,Tsang:2008fd},
and neutron-star
structure~\cite{Horowitz:2000xj,Horowitz:2001ya,Horowitz:2002mb,Carriere:2002bx,Steiner:2004fi,Li:2005sr,Lattimer:2006xb,Fattoyev:2010tb}.
Conversely, due to major advances in all these areas significant constraints on $L$ 
are starting to emerge as one combines theoretical, experimental, and observational 
information\,\cite{Piekarewicz:2007dx,Carbone:2010az,Hebeler:2010jx,Steiner:2011ft,Hebeler:2013nza,Tsang:2012se,Lattimer:2012nd}. 
It is the aim of this special volume to summarize the successes of the past and the 
challenges of the future in understanding the density dependence of the symmetry
energy. In particular, this review contributes to this common goal by using powerful
insights from nuclear collective excitations.

The manuscript has been organized as follows. In Sec.\,\ref{formalism} we review 
the relativistic random phase approximation (RPA) formalism required to compute 
the distribution of isoscalar monopole and isovector dipole strength. Particular 
emphasis is placed on the role of these resonances in constraining the density 
dependence of the symmetry energy. In Sec.\,\ref{results} we present results from
a variety of nuclear energy density functionals (EDFs) and highlight the impact of 
these collective modes in constraining these models. Finally, we offer 
our conclusions and summarize the challenges for the future in Sec.\,\ref{conclusions}.

\section{Formalism}
\label{formalism}

This section is subdivided into three subsections. In the first subsection we introduce the relativistic 
formalism required to compute the distribution of both monopole and dipole strength. Then, we proceed 
to discuss the merits of the distribution of isoscalar monopole strength, particularly in neutron-rich 
nuclei, in constraining simultaneously the incompressibility coefficient of symmetric nuclear matter 
and the density dependence of the symmetry energy. Finally, the last subsection is reserved to a 
discussion of the isovector dipole resonance. In particular, we identify the electric dipole polarizability
as a strong isovector indicator.

\subsection{Relativistic Density Functional Theory}
\label{relformalism}

Historically, relativistic models of nuclear structure were strictly limited to renormalizable
field theories\,\cite{Walecka:1974qa,Serot:1979dc}. The appeal of renormalizability was 
that a handful of model parameters could be calibrated to well-known physical observables 
so that one could then later extrapolate to unknown physical regions without the need for 
introducing additional parameters. However, the modern viewpoint suggests that these 
relativistic models should be treated as effective field theories (EFTs) where the demand 
for renormalizability is no longer required. Yet, EFTs should continue to provide a consistent 
framework for the nuclear many-body problem\,\cite{Furnstahl:2000in}. An effective field 
theory is designed to describe low-energy physics without any attempt at accounting for 
its detailed short-distance behavior. By calibrating the model directly to physical observables, 
the short-distance structure of the theory, as well as other complicated many-body effects, 
gets implicitly encoded in the parameters of the model. In this regard, density functional 
theory (DFT) provides a powerful framework for the construction of an effective field 
theory\,\cite{Hohenberg:1964zz,Kohn:1965,Kohn:1999}, A meaningful criterion used to 
construct the relativistic density functional was proposed by Furnstahl and collaborators 
based on the concept of {\sl ``naive dimensional analysis''} and
{\sl ``naturalness''}~\cite{Furnstahl:1996wv,Furnstahl:1996zm,Rusnak:1997dj,Furnstahl:1997hq,Kortelainen:2010dt}. 
Such an approach is
both useful and powerful as it allows an organizational scheme based
on an expansion in powers of the meson fields; terms in the effective Lagrangian 
with a large number of meson fields are suppressed by a large mass scale. In principle 
then, all terms to a given order must be retained. In practice, 
however, many of these terms have been be ignored. The ``justification'' behind these fairly 
ad-hoc procedure is that whereas the neglected terms are of the same order in a 
power-counting scheme, the full set of parameters is poorly constrained by existing 
experimental data. Thus, ignoring a subset of these terms does not compromise the 
quality of the fit\,\cite{Furnstahl:1996wv,Mueller:1996pm}. 

\subsubsection{Relativistic Lagrangian}
\label{rlagrangian}

The starting point for the relativistic calculation of the nuclear response is the
interacting Lagrangian density of Ref.\,\cite{Mueller:1996pm} supplemented by 
an isoscalar-isovector term originally introduced in Ref.\,\cite{Horowitz:2000xj}. 
That is,
\begin{eqnarray}
 {\mathscr L}_{\rm int} &=&
\bar\psi \left[g_{\rm s}\phi   \!-\! 
         \left(g_{\rm v}V_\mu  \!+\!
    \frac{g_{\rho}}{2}{\mbox{\boldmath $\tau$}}\cdot{\bf b}_{\mu} 
                               \!+\!    
    \frac{e}{2}(1\!+\!\tau_{3})A_{\mu}\right)\gamma^{\mu}
         \right]\psi \nonumber \\
                   &-& 
    \frac{\kappa}{3!} (g_{\rm s}\phi)^3 \!-\!
    \frac{\lambda}{4!}(g_{\rm s}\phi)^4 \!+\!
    \frac{\zeta}{4!}   g_{\rm v}^4(V_{\mu}V^\mu)^2 \nonumber \\
                   &+&     
   \Lambda_{\rm v}\Big(g_{\rho}^{2}\,{\bf b}_{\mu}\cdot{\bf b}^{\mu}\Big)
                           \Big(g_{\rm v}^{2}V_{\nu}V^{\nu}\Big)\;.
 \label{LDensity}
\end{eqnarray}
Motivated by a desire to provide a Lorentz covariant extrapolation to 
dense neutron-star matter, Walecka introduced a Lagrangian density  
with an isodoublet nucleon field $\psi$ interacting through the exchange 
of two massive isoscalar ``mesons": a scalar $\phi$ and a vector 
$V^{\mu}$\,\cite{Walecka:1974qa}. Remarkably, such a simple model 
was already able to account for nuclear saturation at the mean-field level. 
Although the two parameters of the model ($g_{\rm s}$ and $g_{\rm v}$) 
were adjusted to reproduce the density and binding energy at saturation, 
the saturation mechanism was identified as being of relativistic origin. 
Indeed, whereas the vector repulsion continues to increase with baryon density, 
the scalar attraction, which is proportional to the ``Lorentz contracted"
scalar density, saturates. Moreover, to properly describe the equilibrium
density and binding energy, both the scalar and vector mean fields were
found to be very large (of about half of the nucleon rest mass). We note 
that large and canceling scalar and vector fields are the hallmark of the 
relativistic mean-field (RMF) theory. We also note that the pseudoscalar
pion does not contribute at the mean-field level for a ground state of definite 
parity.

In order to move beyond infinite nuclear matter and to be able to describe
the properties of finite nuclei, an isovector meson field ($b^{\mu}$) and the 
photon ($A^{\mu}$) were introduced. Relativistic mean-field equations for 
spherical nuclei were then solved self-consistently and a comparison against
experiment revealed a level of agreement equivalent to that of the most 
sophisticated non-relativistic calculations available at the 
time\,\cite{Horowitz:1981xw,Serot:1984ey}. In particular, a non-relativistic 
reduction of the mean-field equations leads to a Schr\"odinger-like equation
with unique central and spin-orbit potentials. Whereas the central potential 
displays the characteristic cancelation between the strong scalar 
and vector potentials, the strong potentials contribute coherently to the 
spin-orbit potential. Thus, the simple RMF approach was successful in 
accounting for both the relatively weak binding energy and strong spin-orbit 
splitting displayed by the single-particle spectrum. 

In spite of the enormous success of the early RMF models, they all suffered 
from a major shortcoming: the incompressibility coefficient of symmetric 
nuclear matter was predicted to be $K_{0}\!\approx\!550$\,MeV. This was 
recognized early on to be excessively large. Hence, in an effort to overcome
this deficiency, Boguta and Bodmer introduced cubic and quartic scalar meson 
self-interactions \,\cite{Boguta:1977xi}. In particular, these terms (denoted by 
$\kappa$ and $\lambda$) may be adjusted in such a way as to make the 
incompressibility coefficient of symmetric nuclear matter consistent with 
measurements of the distribution of isoscalar monopole strength in medium to heavy 
nuclei\,\cite{Youngblood:1999,Lui:2004wm,Uchida:2003,Uchida:2004bs,Li:2007bp,Li:2010kfa,Patel:2013}.
In fact, one may select suitable values for $\rhoz$, $\epsz$, $K_{0}$, and $M^{\ast}$ (with
the latter denoting the effective nucleon mass at saturation density) and then obtain 
four of the isoscalar parameters of the model ($g_{\rm s}$, $g_{\rm v}$, $\kappa$ and 
$\lambda$) by simply solving a set of four linear simultaneous equations\,\cite{Glendenning:2000}.
Later on, two more parameters were introduced to soften the EOS. In the case of
the isoscalar sector, omega-meson self-interactions (as described by the parameter $\zeta$)  
serve to soften the equation of state of symmetric nuclear at high densities. Indeed, 
M\"uller and Serot found possible to build models with different values of $\zeta$ that 
reproduced the same observed nuclear properties at normal densities but which yield 
maximum neutron star masses that differ by almost one solar mass\,\cite{Mueller:1996pm}. 
Such a finding suggests that observations of massive neutron 
stars\,\cite{Demorest:2010bx,Antoniadis:2013pzd}---rather than laboratory experiment---may 
provide the only meaningful constraint on the high-density component of the 
EOS. Finally, $\Lambda_{\rm v}$ was introduced to modify the poorly constrained 
density dependence of the symmetry energy, while leaving the isoscalar sector 
intact\,\cite{Horowitz:2000xj,Horowitz:2001ya}. Doing so has served to uncover powerful 
correlations between the slope of the symmetry energy $L$ and a host of both laboratory 
and astrophysical observables.

\subsubsection{Relativistic Random Phase Approximation}
\label{rrpa}

Having defined the relativistic density functional one can now proceed to compute 
the linear response of the system to an external perturbation. The first step in a 
consistent mean-field plus RPA (MF+RPA) approach to the nuclear response is 
the calculation of various ground-state properties. This procedure is implemented 
by solving the equations of motion associated with the above Lagrangian density 
in a self-consistent, mean-field approximation\,\cite{Serot:1984ey}. For the various 
meson fields the mean-field approximation consists in solving (non-linear) Klein-Gordon 
equations with the appropriate baryon densities appearing as source terms. 
These baryon densities are computed from the nucleon orbitals that are, in turn, 
obtained from solving the one-body Dirac equation in the presence of scalar and 
time-like vector potentials---which themselves are written in terms of the various
meson fields. This procedure must then be repeated until self-consistency is 
achieved. What emerges from such a self-consistent calculation is a set of 
single-particle energies and corresponding set of Dirac orbitals, and the
self-consistently determined scalar and vector mean-field potentials. 
A detailed implementation of the mean-field procedure may be found in 
Ref.\,\cite{Todd:2003xs}. 

To compute the distribution of both monopole and dipole strength it is sufficient 
to concentrate on the longitudinal nuclear response that is defined as
\begin{align}
  S(q,\omega)\!=\!\sum_{n}\Big|\langle\Psi_{n}|\hat{\rho}({\bf q})|
  \Psi_{0}\rangle\Big|^{2}\!\delta(\omega\!-\!\omega_{n})\!=\!-
  \frac{1}{\pi}\Im\Pi({\bf q},{\bf q};\omega) \;,
  \label{SLong}
\end{align}
where $\Psi_{0}$ is the exact nuclear ground state, $\Psi_{n}$ is an excited state 
with excitation energy $\omega_{n}\!=\!E_{n}\!-\!E_{0}$, $\hat{\rho}({\bf q})$ is the 
transition operator, and $\Pi({\bf q},{\bf q};\omega)$ the associated polarization 
tensor. The transition operator is the Fourier transform of the vector density and 
it takes the following form:
\begin{equation}
  \hat{\rho}_{a}({\bf q}) \!=\! \int d^{3}r \, \bar{\psi}({\bf r}) 
  e^{-i{\bf q}\cdot{\bf r}} \gamma^{0} \tau_{a}\psi({\bf r}) \;.
 \label{Rhoq}
\end{equation}
Here $\gamma^{0}\!=\!{\rm diag}(1,1,-1,-1)$ is the zeroth 
component of the Dirac matrices, $\tau_{0}$ is the identity matrix in 
isospin space, and $\tau_{3}\!=\!{\rm diag}(1,-1)$ is the third isospin
matrix. The cornerstone of our theoretical approach is the polarization 
tensor defined in terms of a time-ordered product of two vector
densities. That is,
\begin{align}
  i\Pi_{ab}(x,y) &=  \langle \Psi_{0}| 
  T \Big[\hat{\rho}_{a}(x)\hat{\rho}_{b}(y)\Big] |\Psi_{0}\rangle \\
   &= \int_{-\infty}^{\infty} 
  \frac{d\omega}{2\pi}e^{-i\omega(x^{0}-y^{0})}
  \Pi_{ab}({\bf x},{\bf y};\omega) \;.
\label{Piab}
\end{align}
Connecting the nuclear response to the polarization tensor is highly
appealing as one can bring to bear the full power of the many-body 
formalism into the calculation of an experimental 
observable\,\cite{Fetter:1971,Dickhoff:2005}. In particular, the polarization 
tensor contains all dynamical information related to the excitation spectrum
of the system. Moreover, the spectral content of the polarization tensor is 
both simple and illuminating:  $\Pi_{ab}({\bf x},{\bf y};\omega)$ is an
analytic function of $\omega$---except for simple poles located at the 
excitation energies and with the transition form-factors extracted from 
the residue at the corresponding pole.

The calculation of the nuclear response at the RPA level has the uncorrelated 
(or mean-field) polarization tensor as one of its main ingredients. In a 
mean-field approximation, the polarization tensor may be written exclusively 
in terms of the nucleon mean-field propagator $G_{F}$. That is,
\begin{align}
  & \Pi_{ab}({\bf x},{\bf y};\omega) = \nonumber \\
  & \; \sum_{0<n<F} 
  \overline{U}_{n}({\bf x})\gamma^{0}\tau_{a}\,
    G_{F}\Big({\bf x},{\bf y};+\omega\!+\!E_{n}^{(+)}\Big) 
    \gamma^{0}\tau_{b}U_{n}({\bf y}) \,+ \nonumber \\
 & \; \sum_{0<n<F} 
    \overline{U}_{n}({\bf y})\gamma^{0}\tau_{b}\,
    G_{F}\Big({\bf y},{\bf x};-\omega\!+\!E_{n}^{(+)}\Big) 
    \gamma^{0}\tau_{a}U_{n}({\bf x}) \;,
\label{pifddf}
\end{align}
where $U_{n}({\bf x})$ and $E_{n}^{(+)}$ are single-particle orbitals and energies 
determined from the self-consistent procedure, and the sum is limited to 
occupied positive-energy orbitals ({\sl i.e.,} below the Fermi surface). The Feynman
mean-field propagator admits a spectral decomposition in terms of the 
mean-field solutions to the Dirac equation, namely,
\begin{equation}
  G_{F}({\bf x},{\bf y};\omega) \!=\! \sum_{n}\!
   \left[
     \frac{U_{n}({\bf x})\overline{U}_{n}({\bf y})}
          {\omega - E_{n}^{(+)} + i\eta} \!+\! 
     \frac{V_{n}({\bf x})\overline{V}_{n}({\bf y})}
          {\omega + E_{n}^{(-)} - i\eta}\!  
   \right].
 \label{gfeyn}
\end{equation}
Here $U_{n}$ and $V_{n}$ are the positive- and negative-energy solutions to the 
Dirac equation with the sum now over all states in the spectrum. Self-consistency
demands that the nucleon propagator must satisfy an inhomogeneous Dirac equation
with a mean-field potential identical to the one used to generate the ground state. 
Only then can fundamental symmetries be maintained, such as the conservation 
of the vector current and the decoupling of the spurious state from the RPA
response. 

As alluded earlier, the analytic structure of the polarization tensor provides critical 
insights. Indeed, the polarization tensor is an analytic function of $\omega$, except 
for simple poles at $\omega\!=\!E_{n}^{(+)}\!-\!E_{m}^{(+)}\!>\!0$, 
where $E_{n}^{(+)}(E_{m}^{(+)})$ is a positive-energy orbital above(below) the Fermi 
surface. Moreover, the corresponding
residues at the pole yield the transition form factors. Finally, note that the contribution 
from the negative-energy orbitals is purely real, as it is free of singularities for 
$\omega\!>\!0$. This implies that negative-energy states make no contribution
to the mean-field response. However, they play an essential role in the RPA 
response---as they are instrumental in ensuring current conservation and the
decoupling of the spurious state\,\cite{Dawson:1990wp}. Thus, in addition to the 
conventional particle-hole excitations, consistency demands the inclusion of pairs 
formed from occupied positive- and negative-energy states. Note that by
themselves, the positive-energy states are not complete. We display in 
Fig.\,\ref{Fig1} the singularity structure of the nucleon propagator at finite
density.
\begin{figure}[ht]
 \vspace{-0.3cm}
 \begin{center}
  \includegraphics[width=0.90\linewidth,angle=0]{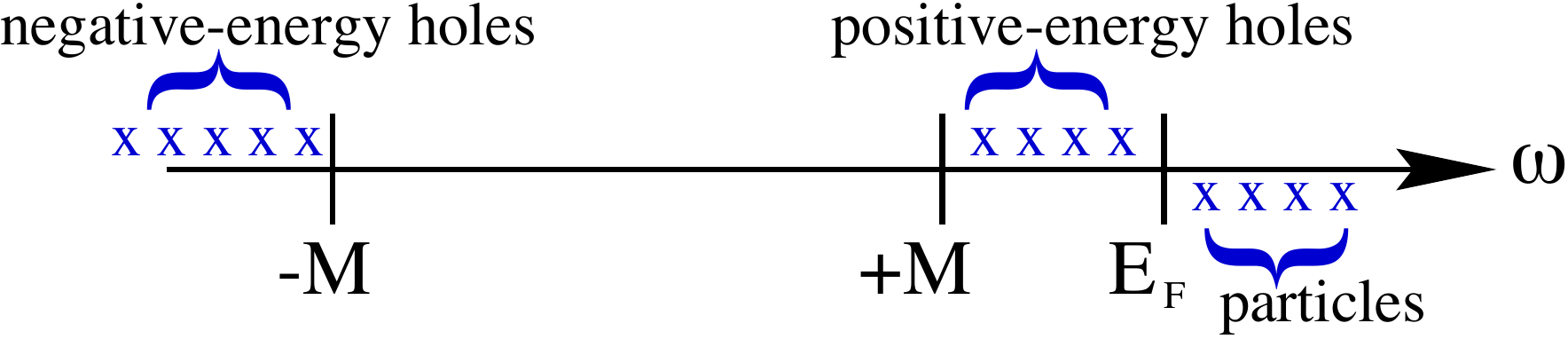}
  \caption{(Color online) Analytic structure of the mean-field nucleon propagator. The propagator
  has singularities at the corresponding mean-field energies. For occupied orbitals---including the
  negative-energy sea---the singularities are located above the real axis; for empty states above
  the Fermi surface they are located below. In addition to the conventional particle-hole excitations,
  consistency demands transitions involving negative-energy states and occupied positive-energy
  orbitals\,\cite{Dawson:1990wp}.}
 \label{Fig1}
 \end{center} 
 \vspace{-0.75cm}
\end{figure}

Although a spectral approach provides valuable physical insights, a scheme that efficiently incorporates 
all the required symmetries---while avoiding any reliance on artificial cut-offs and truncations---is the 
nonspectral approach\,\cite{Piekarewicz:2000nm,Piekarewicz:2001nm}. A nonspectral approach has the 
virtue that both positive- and negative-energy states (bound and continuum) are treated on equal footing. 
In this way the continuum width is treated exactly in the model. Thus, to obtain the mean-field 
propagator in nonspectral form one must solve the Green's problem for the Dirac equation in the presence of 
a mean-field potential identical to the one used to generate the mean-field ground state.

Having generated the lowest-order (mean-field) polarization one may now proceed to compute the fully 
correlated RPA response. One goes beyond the single-particle response by building collectivity into the 
nuclear response through the mixing of many particle-hole excitations. This procedure is implemented by 
iterating the lowest-order polarization to all orders; see Fig.\,\ref{Fig2} for a diagrammatic representation. 
Given that the iteration is to all orders, the analytic structure of the propagator---and thus the location of 
the singularities---is modified relative to the lowest-order predictions. If many pairs are involved, then the 
nuclear response is strongly collective and one ``giant resonance" dominates, namely, it exhausts most 
of the classical sum rule. Such an iterative procedure yields Dyson's equation for the RPA polarization whose 
solution embodies the collective response of the ground state\,\cite{Fetter:1971,Dickhoff:2005}. That is,
\begin{align}
  & \Pi_{ab}^{\rm RPA}({\bf q},{\bf q}';\omega)  =
     \Pi_{ab}({\bf q},{\bf q}';\omega) + \nonumber \\
  & \int\!\frac{d^3k}{(2\pi)^{3}}\frac{d^3k'}{(2\pi)^{3}}
  \Pi_{ac}({\bf q},{\bf k};\omega)           
   V_{cd}({\bf k},{\bf k}';\omega)
  \Pi_{db}^{\rm RPA}({\bf k}',{\bf q}';\omega) \;,
 \label{PiabRPA} 
\end{align}
where $V_{cd}({\bf k},{\bf k}';\omega)$ is the residual interaction (see below) and 
$\Pi_{ab}({\bf q},{\bf q}';\omega)$ is the Fourier transform of the lowest-order polarization:
\begin{equation}
  \Pi_{ab}({\bf q},{\bf q}';\omega)\!=\!\!\int{d^{3}x}\,{d^{3}y}\, 
  e^{-i({\bf q} \cdot{\bf x}-{\bf q}'\cdot{\bf y})}\,
  \Pi_{ab}({\bf x},{\bf y};\omega) \;.
 \label{piqq}
\end{equation}
\begin{figure}[ht]
 \vspace{-0.3cm}
 \begin{center}
  \includegraphics[width=0.90\linewidth,angle=0]{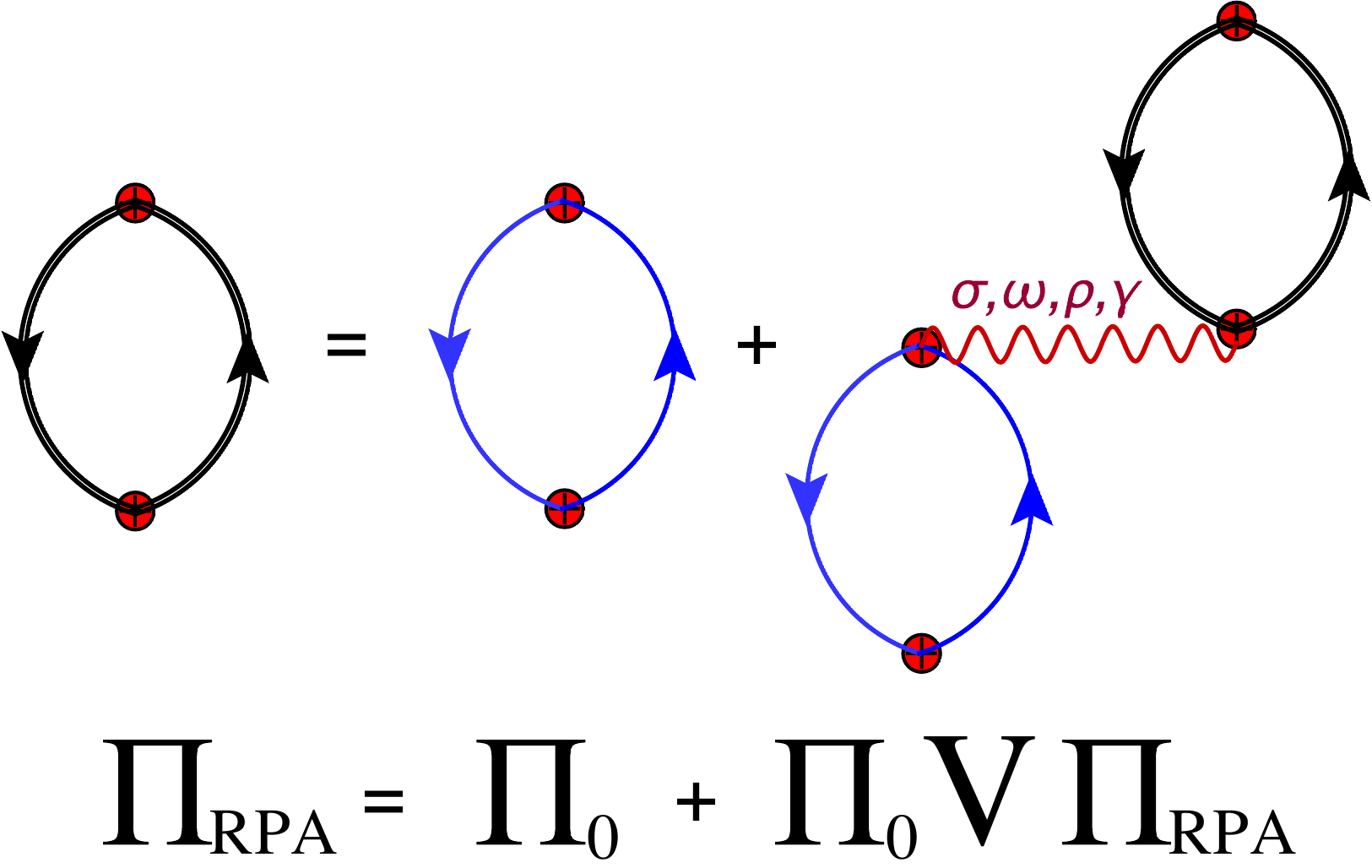}
  \caption{(Color online) Diagrammatic representation of the RPA equations. The ring with the thick
  black lines represents the fully correlated RPA polarization while the one depicted with the thin 
  blue lines is the uncorrelated (mean-field) polarization. The residual interaction denoted with the
  red wavy line is identical to the one used to generate the mean-field ground state.}
 \label{Fig2}
 \end{center} 
 \vspace{-0.75cm}
\end{figure}

The computational demands imposed on relativistic RPA calculations are enormous. Powerful 
symmetries that were present in infinite nuclear matter, such as translational invariance, are now 
broken in the finite system. As a result, the RPA equations that were algebraic in the infinite system 
become integral equations in the finite nucleus. Moreover, modes of excitation that were uncoupled 
before (such as longitudinal-transverse or isoscalar-isovector) become coupled in the finite system. 
Indeed, for nuclei with large neutron excess, the mixing of isoscalar and isovector modes is strong.
Further, because of the ubiquitous meson mixing in relativistic theories (for example, scalar-vector 
mixing) the RPA equations become a complicated set of $9\!\times\!9$ coupled integral equations 
(1 involving $\phi$, 4 involving $V^{\mu}$, and 4 involving $b^{\mu}_{3}$; the photon, being 
``half isoscalar and half isovector'', may be absorbed into the last two terms). Finally, note that because
of meson self-interactions, their respective propagators are no longer local in momentum space and
depart from their simple Yukawa form. For example, the scalar propagator, given by 
$V_{ss}({\bf x},{\bf y}; \omega)\!=\!g_{\rm s}^{2}\Delta_{\rm s}({\bf x},{\bf y}; \omega)$, satisfies the 
following complicated Klein-Gordon equation\,\cite{Piekarewicz:2001nm,Ma:2001hv}: 
\begin{equation}
  \Big(\omega^{2}\!+\!\nabla^{2}\!-m_{s}^{2}\!-\!\kappa\phi\!-\!\frac{1}{2}\lambda\phi^{2}\Big)
  \Delta_{s}({\bf x},{\bf y}; \omega)\!=\!\delta({\bf x}\!-\!{\bf y})\;.
 \label{KG}
\end{equation}
Finally, with the mean-field polarization tensor $\Pi_{ab}$ and the residual interaction $V_{ab}$ 
at hand, one can now solve Dyson's equation for the RPA polarization tensor using matrix inversion 
techniques. Note that whereas excitation modes become coupled in the finite
system, the total angular momentum $J$ and parity $\pi$ remain good quantum numbers. Thus, 
the polarization tensor is decomposed into its various $J^{\pi}$ components and one solves the
RPA equations for only the channels of interest; $J^{\pi}\!=\!0^{+}$ and $J^{\pi}\!=\!1^{-}$ 
in our particular case. 

\subsection{Isoscalar Giant Monopole Resonance}
\label{isgmr}

We start this section by relaxing the early assumption of an incompressible liquid drop to allow 
for density fluctuations around the equilibrium density. In general, the energy per particle of 
asymmetric nuclear matter---at zero temperature and relative to the nucleon mass $M$---may 
be written as follows:
\begin{equation}
  \frac{E}{A}(\rho,\alpha) -\!M \equiv {\cal E}(\rho,\alpha)
                          = {\cal E}_{\rm SNM}(\rho)
                          + \alpha^{2}{\cal S}(\rho)  
                          + {\cal O}(\alpha^{4}) \,.
 \label{EOS}
\end {equation}
Given that the neutron-proton asymmetry is constrained to the interval $0\!\le\alpha\le\!1$, the total 
energy per particle ${\cal E}(\rho,\alpha)$ is customarily expanded in a power series in $\alpha^{2}$. 
Note that no odd powers of $\alpha$ appear as the nuclear force is isospin symmetric. The leading 
term in this expansion ${\cal E}_{\rm SNM}(\rho)$ represents the energy of symmetric nuclear matter. 
In turn, the first-order correction to the symmetric limit is the symmetry energy ${\cal S}(\rho)$. To a 
very good approximation, the symmetry energy measures the energy cost in converting symmetric 
nuclear matter into pure neutron matter. That is,
\begin{equation}
 {\cal S}(\rho)\!\approx\!{\cal E}(\rho,\alpha\!=\!1) \!-\!  {\cal E}(\rho,\alpha\!=\!0) \;.
 \label{SymmE}
\end {equation}
In spite of the enormous progress made over the last years in measuring the masses of exotic 
nuclei\,\cite{Audi:2002rp}, gaining access to the symmetry energy is hindered by the small
$\alpha^{2}$ prefactor. For example, even for ${}^{132}$Sn---a nucleus with a significant neutron 
excess---the neutron-proton asymmetry amounts to only $\alpha^{2}\!=\!(0.242)^{2}\!=\!0.059$. 
And whereas nuclear masses have been measured accurately enough to provide valuable 
constraints on the symmetry energy at a density of $\rho\!\approx\!0.1\,{\rm fm}^{-3}$, its 
density dependence remains largely undetermined\,\cite{Furnstahl:2001un}. 

However, the response of the nuclear ground-state to density fluctuations is sensitive to 
the density dependence of the symmetry energy. To elucidate this fact, we expand both the 
energy of symmetric nuclear matter and the symmetry energy around the equilibrium density. 
That is\,\cite{Piekarewicz:2008nh},
\begin{subequations}
\begin{align}
 & {\cal E}_{\rm SNM}(\rho) = \epsz + \frac{1}{2}K_{0}x^{2}+\frac{1}{6}Q_{0}x^{3} +\ldots \\
 & {\cal S}(\rho) = J + Lx + \frac{1}{2}K_{\rm sym}x^{2}+\frac{1}{6}Q_{\rm sym}x^{3} +\ldots   
\end{align} 
\label{EandS}
\end{subequations}
where $x\!=\!(\rho-\rhoz)\!/3\rhoz$ is a dimensionless parameter that quantifies the deviations 
of the density from its value at saturation. Moreover, $K_{0}$ and $Q_{0}$ are the incompressibility 
coefficient and skewness parameter of symmetric nuclear matter; $K_{\rm sym}$ and $Q_{\rm sym}$ 
are the corresponding quantities for the symmetry energy. Note that unlike symmetric nuclear matter, 
the linear term in $x$ (namely $L$) does not vanish in the case of the symmetry energy. By relying 
on these expansions, the energy per particle of asymmetric nuclear matter may be written in the 
following form [{\sl c.f.} to Eq.\,(\ref{EAsym})]:
\begin{align}
 {\cal E}(\rho,\alpha) &= (\epsz\!+\!J\alpha^{2}) +L\alpha^{2}x 
                                + \frac{1}{2}(K_{0}\!+\!\alpha^{2}K_{\rm sym})x^{2} \nonumber \\
                              &+ \frac{1}{6}(Q_{0}\!+\!\alpha^{2}Q_{\rm sym})x^{3} + \ldots
 \label{EvsX}
\end{align}
Since the density pressure $L$ does not vanish, the saturation point in asymmetric
matter shifts from $x_{0}\!=\!0$ to $\bar{x}_{0}$, where the latter is defined as the 
solution to $\partial{\cal E}/{\partial x}\!=\!0$. To ${\cal O}(\alpha^{2})$, the solution 
to this equation is simple and results in a shift of the saturation density 
to\,\cite{Piekarewicz:2008nh}:
\begin{equation}
  \bar{x}_{{}_{0}}=-\frac{L}{K_{0}}\alpha^{2} \implies
  \frac{\barrhoz}{\rhoz} \!=\! 1\!+\!3\bar{x}_{{}_{0}}
   \!=\!1\!-\!3\frac{L}{K_{0}}\alpha^{2}\;.
 \label{X0Bar}
\end{equation}
Given that PREX has established the existence of a neutron rich skin in ${}^{208}$Pb, the 
symmetry pressure must be positive, indicating that the saturation point must move to lower 
densities. In turn, the incompressibility coefficient of neutron-rich matter may now be found 
by expanding Eq.\,(\ref{EvsX}) around $\bar{x}_{0}$. That is,
\begin{equation}
 {\cal E}(\rho,\alpha) = {\cal E}(\bar{x}_{{}_{0}},\alpha)
 +\frac{1}{2}(x\!-\!\bar{x}_{{}_{0}})^{2}
 \left(\frac{\partial^{2}{\cal E}}{\partial x^{2}}\right)_{\bar{x}_{0}}
 +\ldots
 \label{NewEOS}
\end{equation}
Alternatively, by introducing $\bar{x}\!=\!(\rho\!-\!\barrhoz)\!/3\barrhoz$ to quantify 
deviations from the new equilibrium density, we obtain in analogy to the symmetric
case, the following expression:
\begin{equation}
 {\cal E}(\rho,\alpha) = (\epsz+J\alpha^{2}) +\frac{1}{2}K_{0}(\alpha)\bar{x}^{2}
 +\ldots\;
 \label{SNMlike}
\end{equation}
where the incompressibility coefficient of neutron-rich matter is given to
${\cal O}(\alpha^{2})$ by\,\cite{Piekarewicz:2008nh} 
\begin{equation}
  K_{0}(\alpha)=K_{0}+K_{\tau}\alpha^{2}
            \equiv K_{0}+\Big(K_{\rm sym}-6L-\!\frac{Q_{0}}{K_{0}}L\Big)\alpha^{2} \;.
\label{Kalpha}            
\end{equation}
Given that the centroid energy of the giant monopole resonance (GMR)
scales as the square-root of the incompressibility coefficient\,\cite{Blaizot:1995}, 
the impact of the density dependence of the symmetry energy on the distribution 
of monopole strength may be significant. In particular, a systematic study of GMR 
energies as a function of neutron-proton asymmetry---especially along isotopic 
chains\,\cite{Li:2007bp,Li:2010kfa,Patel:2013}---may be of enormous value. For 
example ${}^{90}$Zr, a nucleus with a well-developed GMR peak but with a 
neutron-proton asymmetry of only $\alpha_{90}^{2}\!=\!(0.111)^{2}\!=\!0.012$, is 
ideal for determining the incompressibility coefficient of symmetric nuclear matter 
$K_{0}$. Once $K_{0}$ has been determined, one may then use the GMR in 
${}^{208}$Pb, with a value of $\alpha_{208}^{2}\!=\!(0.212)^{2}\!=\!0.045$ four times 
as large, to constrain $K_{\tau}$\,\cite{Piekarewicz:2003br}. It is important to underscore 
that one measurement alone is not sufficient. For example, NL3 is 
known to predict a rapid increase with density for both symmetric nuclear matter 
and the symmetry energy. In particular, it was possible for NL3 to be consistent 
with the GMR in ${}^{208}$Pb by having a large value of $K_{\tau}$ compensate 
for a correspondingly large value of $K_{0}$\,\cite{Piekarewicz:2002jd}. However,
this was no longer possible in the case of ${}^{90}$Zr because its significant lower 
value of $\alpha^{2}$. Thus, NL3 was found to overestimate the GMR in ${}^{90}$Zr 
by about 0.7\,MeV, which is significantly larger than the 0.2\,MeV experimental 
error\,\cite{Piekarewicz:2003br}.

We conclude this section with a brief summary of the relevant moments of the 
distribution of strength that will be displayed in Sec.\,\ref{results}. In the long 
wavelength limit the distribution of isoscalar monopole 
strength $R(\omega;E0)$ is directly related to the longitudinal response defined in
Eq.\,(\ref{SLong}) by the following expression:
\begin{equation}
  R(\omega;E0) = 
  \lim_{q\rightarrow 0} \left(\frac{36}{q^{4}}\right) S_{L}(q,\omega;E0) \;. 
  \label{RGMR}
\end{equation}
In turn, moments of the distribution of isoscalar monopole strength are defined as 
suitable energy weighted sums. That is,
\begin{equation}    
  m_{n}(E0) \equiv \int_{0}^{\infty}\!\omega^{n} R(\omega;E0)\, d\omega \;. 
 \label{GMRMoments}
\end{equation}
Widely used in the literature are the energy weighted $m_{1}$, energy unweighted $m_{0}$, 
and inverse energy weighted $m_{-1}$ sums. In particular, we will present results in
Sec.\,\ref{results} for the GMR centroid energies defined as $E_{\rm GMR}\!=\!m_{1}/m_{0}$.

\subsection{Isovector Giant Dipole Resonance}
\label{ivgdr}

In a simple macroscopic picture the isovector giant dipole resonance (GDR) is perceived 
as an out of phase collective oscillation of neutrons against protons. Thus, the symmetry 
energy at sub-saturation densities acts as the restoring force. As in the case of the GMR,
the distribution of isovector dipole strength $R(\omega;E1)$ may be extracted from the
longitudinal response defined in Eq.\,(\ref{SLong}). That is, in the long wavelength 
approximation we obtain 
\begin{equation}
 R(\omega;E1) = \lim_{q\rightarrow 0} \left(\frac{9}{4\pi q^{2}}\right)S_{L}(q,\omega;E1)\;.
 \label{RGDR} 
\end{equation}
We note that the relevant transition operator extracted from Eq.\,(\ref{Rhoq}) for the isovector 
dipole mode is given by
\begin{equation}
  \hat{\rho}_{{}_{3}}({\bf q}) \!\propto\!  \int d^{3}r \, \bar{\psi}({\bf r}) j_{1}(qr) Y_{1\mu}({\bf \hat{r}})
  \gamma^{0}\tau_{3}\psi({\bf r}) \;,
 \label{RhoGDR} 
\end{equation}
where $j_{1}$ is a spherical Bessel function and $Y_{1\mu}$ is a spherical harmonic. 
We have mentioned that in finite nuclei, especially those with a significant neutron 
excess, there may be significant mixing between isoscalar and isovector modes. 
In the particular case of dipole resonances, the isovector GDR will mix with the 
isoscalar mode, so care must be taken in handling properly the spurious center
of mass motion. However, we stress that in the context of a self-consistent RPA 
framework, no modification to the transition operator given in Eq.\,(\ref{RhoGDR}) is 
required, as the formalism by itself and nothing else should push all spurious strength 
to zero excitation energy\,\cite{Piekarewicz:2001nm}.

Moments of the distribution of isovector dipole strength may be defined in analogy to 
Eq.\,(\ref{GMRMoments}) as follows:
\begin{equation}    
  m_{n}(E1) \equiv \int_{0}^{\infty}\!\omega^{n} R(\omega;E1)\, d\omega \;. 
 \label{GDRMoments}
\end{equation}
Of great relevance are the energy weighted $m_{1}$ and inverse energy weighted
$m_{-1}$ sums. In particular, $m_{1}$ satisfies a classical energy weighted sum rule 
(EWSR): 
\begin{equation}
 m_{1}(E1) =
 \frac{9\hbar^{2}}{8\pi M}\left(\frac{NZ}{A}\right)\!\approx\!
 14.8 \left(\frac{NZ}{A}\right) {\rm fm}^{2}{\rm MeV} \;.
 \label{EWSR}
\end{equation}
Note that the EWSR is related to the total photoabsorption cross section $\sigma(\omega)$ 
and the corresponding TRK sum rule. That is,
\begin{equation}
 \int_{0}^{\infty}\!\!\sigma(\omega)\,d\omega =
 \frac{16\pi^{3}}{9} \frac{e^{2}}{\hbar c} m_{1} 
 \approx 60 \left(\frac{NZ}{A}\right) {\rm MeV \, mb}\;.
\label{TRK}
\end{equation}
The power of the sum rules is that in principle they are independent of the details of the 
interaction. In practice, however, the classical sum rules are only valid in the absence of 
exchange and momentum dependent forces\,\cite{Harakeh:2001}. The appearance 
of such forces modifies the classical sum rules and their impact is traditionally accounted 
for by multiplying the right-hand side of Eqs.\,(\ref{EWSR}-\ref{TRK}) by a factor of
$(1\!+\!\kappa_{\rm TRK})$ with $\kappa_{\rm TRK}\!\approx\!0.2$\,\cite{Harakeh:2001}. 
Also note that the classical sum rules were derived using a non-relativistic formalism so 
one may also need to correct for relativistic effects; we will assume here that such 
relativistic effects have been also incorporated into $\kappa_{\rm TRK}$.

Besides the EWSR, the inverse energy weighted sum $m_{-1}$ is of critical importance 
because its sensitivity to the density dependence of the symmetry 
energy\,\cite{Reinhard:2010wz,Piekarewicz:2010fa,Piekarewicz:2012pp,Roca-Maza:2013mla}. 
Note that the electric dipole polarizability $\alphad$ is simply related to $m_{-1}$:
\begin{equation}
  \alphad = \frac{8\pi}{9}e^{2}m_{-1}\;.
\label{DipPol}
\end{equation}
Recently, a high-resolution measurement of the electric dipole polarizability in ${}^{208}$Pb 
was performed at the Research Center for Nuclear Physics (RCNP) using polarized proton 
inelastic scattering at forward angles\,\cite{Tamii:2011pv,Poltoratska:2012nf}. The great
virtue of this experiment is that at forward angles Coulomb excitation dominates, thereby
making the extraction of $\alphad$ free from strong-interaction uncertainties. The reported
value of $\alphad$ in ${}^{208}$Pb is\,\cite{Tamii:2011pv,Poltoratska:2012nf}
\begin{equation}
  \alphad^{208} = (20.1 \pm 0.6)\, {\rm fm}^{3} \;.
\label{DipPol208}
\end{equation}
For additional details on this landmark experiment and on how it may be used to constrain 
the density dependence of the symmetry energy, see the contribution to this volume by 
Tamii, von\,Neumann-Cosel, and Poltoratska.

Although microscopic MF+RPA calculations of $\alphad$ will be presented in the next
section, we close this section by providing insights from the macroscopic liquid droplet 
model to elucidate the connection between the dipole polarizability and the density 
dependence of the symmetry energy\,\cite{Roca-Maza:2013mla,Satula:2005hy}. In particular, 
following closely the recent analysis by Roca-Maza and collaborators, one finds a simple 
relation between $\alphad$ and two fundamental parameters of the symmetry energy, 
namely, $J$ and $L$. That is\,\cite{Roca-Maza:2013mla},
\begin{equation}
\alphadDM \!\approx\! \frac{\pi e^{2}}{54} \frac{A \langle r^2\rangle}{J}
\left[1+\frac{5}{3}\frac{L}{J}\epsilon_{{}_{A}}\right]\;.
\label{dpdm}
\end{equation}
where $\langle r^2\rangle$ is the mean-square radius of the nucleus and 
$\epsilon_{{}_{A}}\!=\!(\rhoz\!-\!\rho_{{}_{A}})/3\rhoz$ accounts for the difference between 
the saturation density $\rhoz$ and an appropriate average nuclear density $\rho_{{}_{A}}$.
Note that this expression suggests that $J$ and $L$ may be separately constrained from 
measuring $\alphad$ in a few nuclei. Moreover, it indicates that the dipole polarizability 
times the symmetry energy at saturation density ({$\alphad J$) should be better correlated 
to $L$ than $\alphad$ alone\,\cite{Roca-Maza:2013mla}.

\section{Results}
\label{results}
Having developed the formalism required to compute the distribution of both monopole and 
dipole strength, we are now in a position to discuss the predictions of our relativistic RPA 
calculations. However, before doing so it is convenient to introduce the models used in this 
work and their associated predictions for the bulk parameters of infinite nuclear matter. We 
start by displaying in Table\,\ref{Table1} the model parameters for the various relativistic 
density functionals used in this review. Note that the parameters are defined according to 
the Lagrangian density given in Eq.\,(\ref{LDensity}). Included in this set are the two 
accuartely-calibrated density functionals: NL3\,\cite{Lalazissis:1996rd,Lalazissis:1999} and 
FSUGold (or ``FSU'' for short)\,\cite{Todd-Rutel:2005fa}. Relative to NL3---a model enormously 
successful in reproducing masses and charge radii over the whole nuclear chart---FSUGold 
includes two additional parameters ($\zeta$ and $\Lambda_{\rm v}$) that are used to soften 
both the EOS of symmetric nuclear matter and the symmetry energy. The IU-FSU effective 
interaction was conceived in response to the recent interpretation of X-ray observations by 
Steiner, Lattimer, and Brown that suggests that FSUGold predicts 
neutron star radii that are too large and a maximum stellar mass that is too 
small\,\cite{Steiner:2010fz}. As seen in Table\,\ref{Table1}, stiffening the EOS of symmetric 
nuclear matter by reducing $\zeta$ and softening the symmetry energy by increasing 
$\Lambda_{\rm v}$ provides a simple and efficient procedure to overcome these 
problems\,\cite{Fattoyev:2010mx}. Finally, the last three effective interactions 
labeled as TAMUC-FSU (or ``TF'' for short) were created in response to the provocative result 
reported by the PREX collaboration, namely, that the neutron-skin thickness of ${}^{208}$Pb 
is significantly larger than predicted by a large set of theoretical calculations. Indeed, the 
PREX central value of $\rskin^{208}\!\approx\!0.33$\,fm is particularly intriguing. Hence, the 
TF interactions (calibrated to $\rskin^{208}\!=\!0.25$, $0.30$, and $0.33$\,fm, respectively) 
were created to test whether such a thick neutron-skin in ${}^{208}$Pb is already incompatible 
with laboratory experiments or astrophysical observations\,\cite{Fattoyev:2013yaa}. 
\begin{table*}
\begin{center}
\begin{tabular}{|l||c|r|r|r|r|r|c|c|}
 \hline\rule{0pt}{2.25ex} 
 $\!\!$Model & $m_{\rm s}(\rm MeV)$  & $\hfill g_{\rm s}^{2} \hfill$ & $\hfill g_{\rm v}^2 \hfill $ & $\hfill g_{\rho}^2 \hfill $
           & $\hfill  \kappa(\rm MeV) \hfill$  & $\hfill \lambda \hfill $ & $\zeta$ & $\Lambda_{\rm v}$\\
 \hline
 \hline
 NL3                  & 508.194 & 104.3871  & 165.5854 &  79.6000 & 3.8599   &  $-$0.015910 & 0.00 & 0.00000 \\
 FSUGold          & 491.500 & 112.1996   & 204.5469 & 138.4701 & 1.4203  & $+$0.023760 & 0.06 & 0.03000 \\           
 IU-FSU             & 491.500 &   99.4266  & 169.8349 & 184.6877 & 3.3808  & $+$0.000296 & 0.03 & 0.04600 \\
 TAMUC-FSUa   & 502.200 &  106.5045 & 176.1780 &   97.3556 & 3.1824  &  $-$0.003470 & 0.02 & 0.01267 \\
 TAMUC-FSUb   & 497.100 &  104.3524 & 176.2026 &   90.5000 & 3.1163  &  $-$0.003021 & 0.02 & 0.00000 \\  
 TAMUC-FSUc   & 496.800 &  113.9565 & 198.0546 & 103.4000 & 2.6079  &  $-$0.001864 & 0.02 & 0.00000 \\          
 \hline
\end{tabular}
\caption{Model parameters for all the relativistic density functionals used in the text. Masses for the isoscalar-vector 
meson, isovector-vector meson, and nucleon have been fixed at $m_{\rm v}\!=782.5$\,MeV, $m_{\rho}\!=763$\,MeV, 
and  $M\!=\!939$\,MeV, 
respectively.} 
\label{Table1}
\end{center}
\end{table*}

Note that although the parameters of the IU-FSU and TAMUC-FSU models do not follow from a strict optimization procedure, 
a significant effort was made in reproducing some bulk parameters of infinite nuclear matter as well as some critical properties 
of finite nuclei. Indeed, predictions from various bulk parameters of infinite nuclear matter as defined in Eqs.\,(\ref{EandS}) 
and\,(\ref{Kalpha}) are listed in Table\,\ref{Table2}. In particular, the variation among those bulk parameters that are sensitive 
to nuclear masses is relatively small: $\Delta\rhoz\!\approx\!5\%$, $\Delta\epsz\!\approx\!2\%$, and $\Delta J\!\approx\!40\%$. 
Note that even though $\Delta J$ does not appear to be small, Eq.\,(\ref{SNMlike}) suggests that the relevant combination that
enters the determination of nuclear masses is $\alpha^{2}\!\Delta J$; for stable nuclei $\alpha^{2}\!\Delta J\!\lesssim\!6\%$.
Although accurately measured ground-state properties are fairly insensitive to the incompressibility coefficient of symmetric 
nuclear matter, the calibration of modern relativistic functionals now aims to incorporate GMR energies from a few critical 
nuclei, such as ${}^{90}$Zr and ${}^{208}$Pb\,\cite{Chen:2013tca}. This should help reduce the uncertainty in $K_{0}$. Note, 
however, that at present even the sign of the skewness parameter $Q_{0}$ remains unconstrained. Also poorly constrained 
are the symmetry energy parameters $L$ and $K_{\rm sym}$, with the former displaying a variation of more than a factor of 
two and the latter with a sign that still remains undetermined. Not surprisingly, the symmetry energy contribution to the 
incompressibility coefficient $K_{\tau}$---which involves a linear combination of $L$, $K_{\rm sym}$, and $Q_{0}$ as 
shown in Eq.\,(\ref{Kalpha})---displays a variation of more than a factor of four. However, note that in spite of such a 
large variation, the sign of $K_{\tau}$ is predicted to be negative in all the models. This fact is driven by the large and 
negative coefficient (of $-6$) in front of $L$ in the expression for $K_{\tau}$. Given the very strong correlation between 
$L$ and the neutron-skin thickness of ${}^{208}$Pb, $K_{\tau}$ is likely to be negative for all realistic models that predict 
a neutron-rich skin in ${}^{208}$Pb. Finally, the last two columns display the incompressibility coefficient of neutron-rich 
matter with a neutron-proton asymmetry equal to that of ${}^{90}$Zr and ${}^{208}$Pb, respectively. Ignoring the predictions 
from the NL3 parametrization, as it did not include information on GMR energies in its calibration, the variations amount to 
$\Delta K_{90}\!\approx\!10\%$  and $\Delta K_{208}\!\approx\!3\%$. As GMR centroid energies scale as
$E_{\rm GMR}(\alpha)\!\propto\!\sqrt{K_{0}(\alpha)}$, the anticipated errors in the respective energies should be half as 
large. Given that FSUGold is consistent with the GMR centroid energies in both ${}^{90}$Zr and ${}^{208}$Pb, we expect
that all models---with the possible exception of NL3---should also be consistent with experiment. Note that whereas GMR 
constraints on the symmetry energy are hindered by the short lever arm multiplying $K_{\tau}$ ({\sl i.e.,} $\alpha^{2}$), 
isovector dipole excitations probe directly the symmetry energy. In particular, insights from the droplet model suggest 
that measurements of the electric dipole polarizability for a variety of nuclei should simultaneously constrain the value 
and slope ($J$ and $L$) of the symmetry energy at saturation density. Ultimately, the determination of the density 
dependence of the symmetry energy will require a concerted effort involving both laboratory experiments and 
astrophysical observations. This special volume goes a long way towards fulfilling this goal. 

\begin{table*}
\begin{center}
\begin{tabular}{|l||c|c|c|r|c|r|r|c|c|c|}
 \hline\rule{0pt}{2.25ex}
 $\!\!$Model & $\rhoz({\rm fm}^{-3})$ & $\epsz$ & $K_{0}$ & $\hfill Q_{0} \hfill$ & 
 $J$ & $\hfill L \hfill$ & $\hfill K_{\rm sym} \hfill $ & $K_{\tau}$ & $K_{90}$ & $K_{208}$\\
 \hline
 \hline
 NL3                   &  0.148  & $-$16.24 & 271.54 &     209.46 & 37.29 & 118.19 &     100.88 & $-$699.41 & 262.90 & 240.24 \\ 
 FSUGold           &  0.148  & $-$16.30 & 230.00 & $-$522.74 & 32.59 &   60.52 &  $-$51.31 & $-$276.87 & 226.59 & 217.62 \\
 IU-FSU              & 0.155  & $-$16.40 & 231.33 & $-$291.12 & 31.30 &   47.21 &      28.53  & $-$195.29 & 228.92 & 222.59 \\
 TAMUC-FSUa    & 0.149 &  $-$16.23 & 245.00 & $-$150.77 & 35.05 &   82.50 &  $-$68.37 & $-$512.60  & 238.67 & 222.06 \\ 
 TAMUC-FSUb    & 0.149 &  $-$16.40 & 250.00 & $-$156.60 & 40.07 &  122.53 &      45.88 & $-$612.54  & 242.44 & 222.59 \\
 TAMUC-FSUc    & 0.148 &  $-$16.47 & 260.49 &       92.86 & 43.67 &  135.25 &      51.64 & $-$808.04  & 250.51 & 224.33 \\
\hline
\end{tabular}
\caption{Bulk parameters of infinite nuclear matter at saturation density $\rho_{{}_{\hspace{-0.5pt}0}}$. 
The quantities $\varepsilon_{{}_{\hspace{-0.5pt}0}}$, $K_{0}$, and $Q_{0}$ represent the binding energy 
per nucleon, incompressibility coefficient, and skewness parameter of symmetric nuclear matter. 
Similarly, $J$, $L$, and $K_{\rm sym}$  represent the energy, slope, and curvature of the symmetry 
energy; see Eq.\,(\ref{EandS}). $K_{\tau}$ is the symmetry energy contribution to the incompressibility 
coefficient defined in Eq.\,(\ref{Kalpha}). Finally, $K_{90}$ and $K_{208}$ correspond to the incompressibility 
coefficient of asymmetric nuclear matter with a neutron-proton asymmetry equal to that of ${}^{90}Zr$
($\alpha_{{}_{90}}\!=\!0.111$) and ${}^{208}$Pb ($\alpha_{{}_{208}}\!=\!0.211$), respectively. All quantities 
are in MeV unless otherwise indicated.}
\label{Table2}
\end{center}
\end{table*}

\subsection{Isoscalar Giant Monopole Resonance}
\label{res:isgmr}
We start this section by displaying in Fig.\,\ref{Fig3} relativistic RPA predictions for the distribution 
of isoscalar monopole strength for nuclei ranging from ${}^{90}$Zr to ${}^{208}$Pb.
As it was underscored in the previous section, a systematic study on nuclei with a wide range 
of neutron-proton asymmetries and well developed GMR peaks is required to constrain both 
$K_{0}$ and $K_{\tau}$. As it is evident from the figure, the heavier the nucleus the stronger the 
collective effects. Given the strong and attractive nature of the isoscalar residual interaction, 
a softer and better developed GMR peak is generated with increasing $A$. Note that the predicted 
escape width is generated exactly within the non-spectral approach. However, the predicted width 
is in general smaller than experiment as more complicated excitations (such as those involving
many particles and holes) are beyond the scope of the RPA approach.
\begin{figure}[ht]
 \begin{center}
  \includegraphics[width=0.9\linewidth,angle=0]{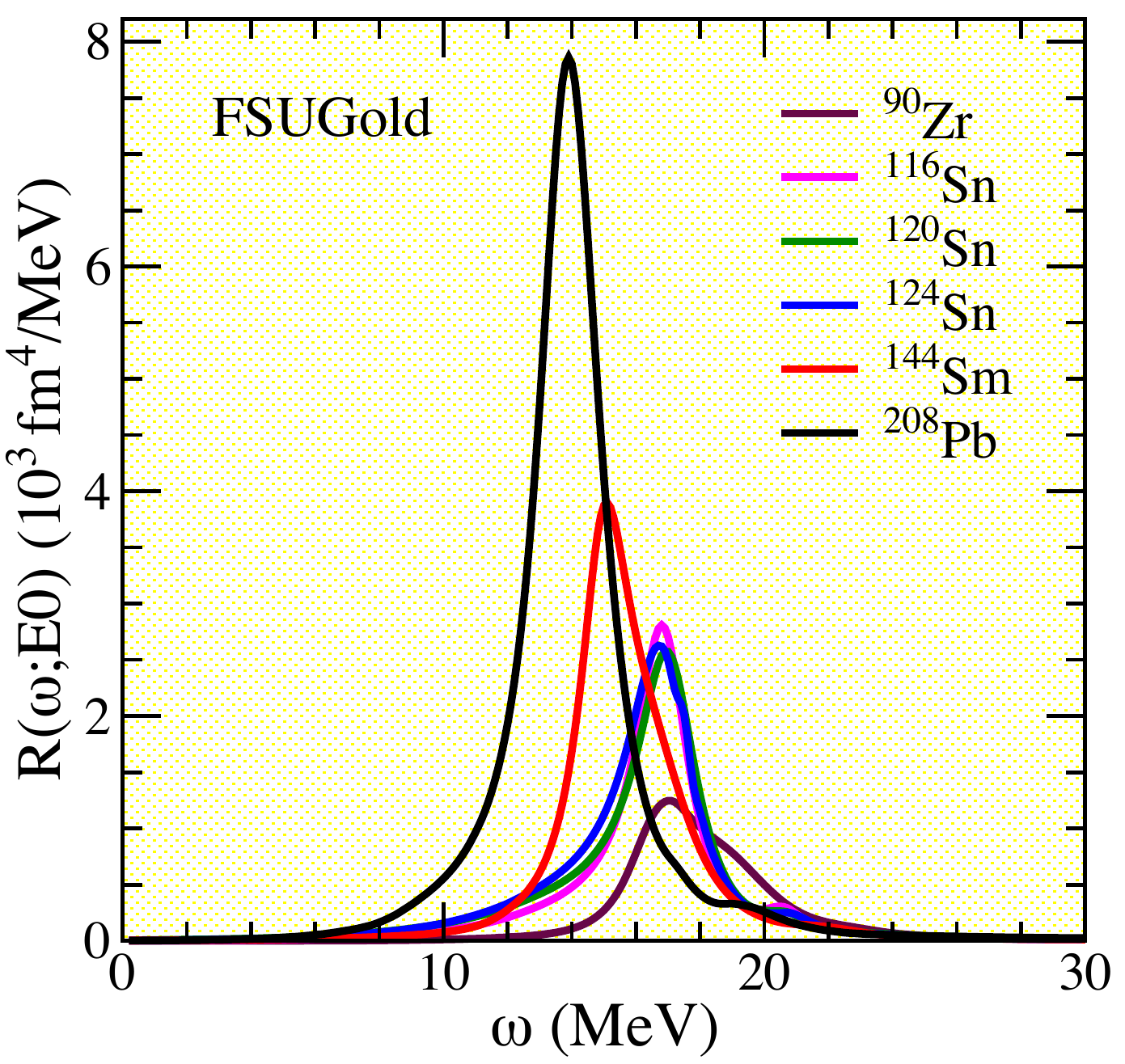}
  \caption{(color online) Distribution of isoscalar monopole strength for nuclei ranging 
  from ${}^{90}$Zr to ${}^{208}$Pb as predicted by relativistic RPA calculations using
  the FSUGold effective interaction\,\cite{Todd-Rutel:2005fa}.}
 \label{Fig3}
 \end{center} 
\end{figure}
Although theoretical RPA calculations are unable to account for the full experimental 
width, they should be able to provide a fairly good description of the centroid energy
$E_{\rm GMR}\!=\!m_{1}/m_{0}$. To this end, predictions for the centroid energy of 
the six nuclei considered in Fig.\,\ref{Fig3} are listed in Table\,\ref{Table3} together 
with the corresponding experimental 
values\,\cite{Youngblood:1999,Lui:2004wm,Uchida:2003,Uchida:2004bs,Li:2007bp,Li:2010kfa,Patel:2013}. 
The numbers in parenthesis correspond to the range of excitation energies used 
to compute the moments of the distribution of strength [see Eq.\,(\ref{GMRMoments})].
The same information is displayed in graphical form in Fig.\,\ref{Fig4} for ${}^{90}$Zr, 
${}^{116}$Sn, ${}^{144}$Sm, and ${}^{208}$Pb. It is evident from the figure that regardless 
of the value of $K_{0}$ and the stiffness of the symmetry energy, all models---with the 
possible exception of NL3---cluster around each other. Indeed, with the exception of NL3, 
the variation in the predictions of these models is limited to at most 2\%. This suggests 
that centroid energies of monopole resonances---even those of nuclei with a large neutron 
excess---are unable to place stringent constrains on the density dependence of the
symmetry energy\,\cite{Fattoyev:2013yaa}. 

\begin{table*}
\begin{center}
\begin{tabular}{|l||c|c|c|c|c|c|}
 \hline\rule{0pt}{2.5ex}
 $\!\!$Model & ${}^{90}$Zr(10-26) & ${}^{116}$Sn(10-23) & ${}^{120}$Sn(10-20) 
                  & ${}^{124}$Sn(10-20) & ${}^{144}$Sm(8-23) & ${}^{208}$Pb(8-23) \\ 
 \hline
 NL3                  & 18.62 & 17.09 & 16.76 & 16.52 & 16.14 & 14.33 \\ 
 FSUGold          & 17.98 & 16.55 & 16.24 & 16.08 & 15.63 & 14.02 \\     
 IU-FSU            & 17.87 & 16.51 & 16.29 & 16.14 & 15.61 & 14.17 \\             
 TAMUC-FSUa  & 18.06 & 16.59 & 16.32 & 16.10 & 15.65 & 13.96 \\
 TAMUC-FSUb  & 18.09 & 16.61 & 16.31 & 16.08 & 15.67 & 13.89 \\
 TAMUC-FSUc  & 18.17 & 16.66 & 16.36 & 16.11 & 15.72 & 13.89 \\
 \hline
 \hline
 Experiment & & & & & & \\
 \hline
 TAMU & 17.89$\pm$0.20 & 16.07$\pm$0.12 &                         &14.50$\pm$0.14 
           & 15.39$\pm$0.28 & 14.17$\pm$0.28 \\  
 RCNP & 18.10$\pm$0.10 & 16.19$\pm$0.10 & 15.55$\pm$0.10 & 15.27$\pm$0.10 
           & 16.00$\pm$0.35 & 13.90$\pm$0.20 \\
\hline
\end{tabular}
\caption{GMR Centroid energies predicted by the relativistic models discussed in the text. 
Quantities in parenthesis represent the minimum ($\omega_{\rm min}$) and maximum
($\omega_{\rm max}$) values used in Eq.\,(\ref{GMRMoments}) to compute various
moments of the distribution of strength. Experimental values were obtained at
TAMU\,\cite{Youngblood:1999,Lui:2004wm} and 
RCNP\,\cite{Uchida:2003,Uchida:2004bs,Li:2007bp,Li:2010kfa,Patel:2013}.}
\label{Table3}
\end{center}
\end{table*}

\begin{figure}[ht]
 \begin{center}
  \includegraphics[width=0.9\linewidth,angle=0]{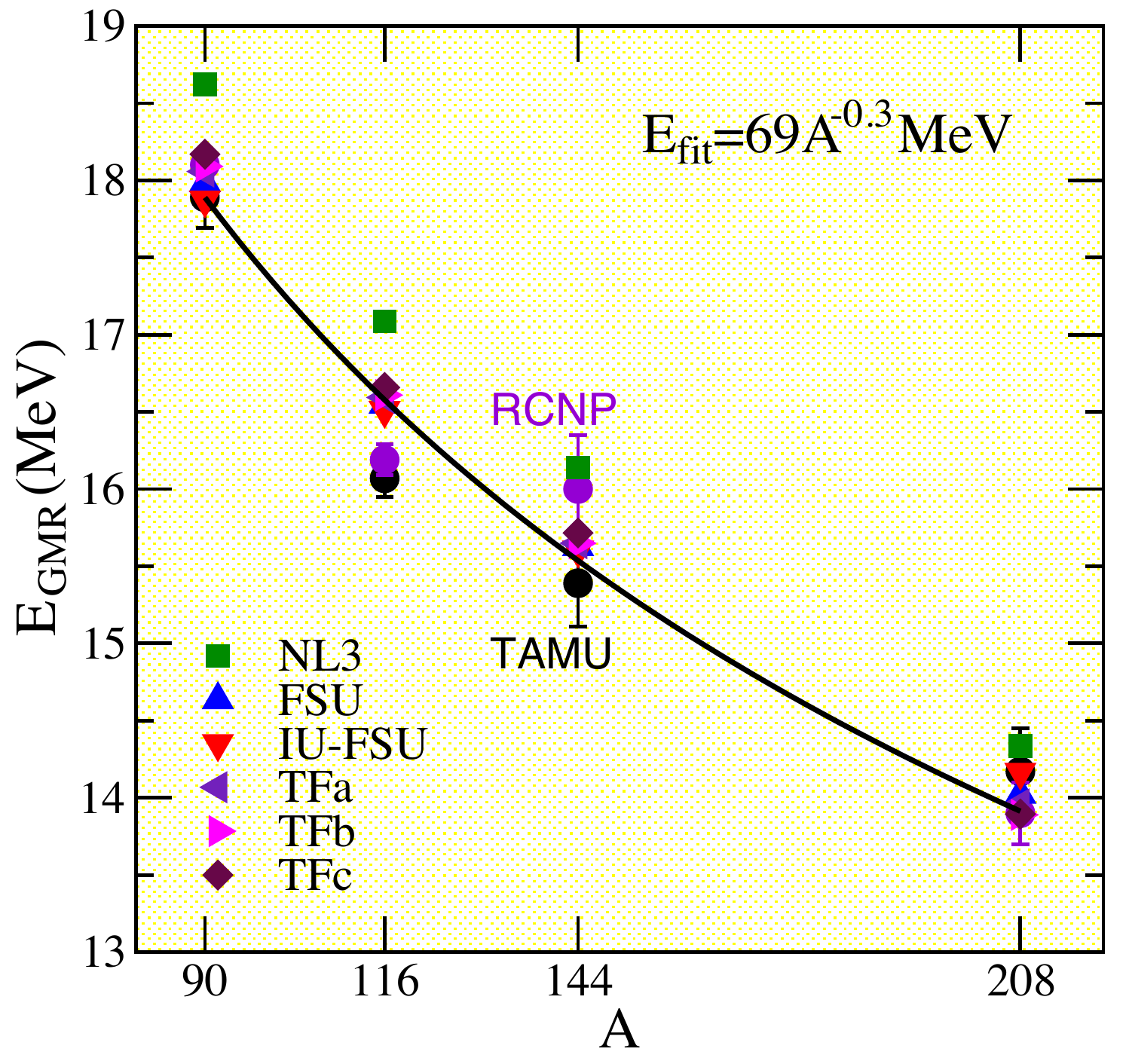}
  \caption{(color online) GMR centroid energies ($E_{\rm GMR}\!=\!m_{1}/m_{0}$) in  
  ${}^{90}$Zr, ${}^{116}$Sn, ${}^{144}$Sm, and ${}^{208}$Pb, as predicted by the
  relativistic effective interactions used in the text. Also shown in the figure are
  experimental measurements reported by TAMU\,\cite{Youngblood:1999,Lui:2004wm}  
  and RCNP\,\cite{Uchida:2003,Uchida:2004bs,Li:2007bp,Li:2010kfa,Patel:2013}. The
  black line represents a best fit to the FSU predictions of the form
  $E_{\rm fit}\!=\!69\,{\rm A}^{-0.3}$.} 
 \label{Fig4}
 \end{center} 
\end{figure}

\subsubsection{Why is Tin so soft?}
\label{SoftTin}

We just showed how predictions from a variety of models with widely different bulk properties can 
reproduce the GMR centroid energies of ${}^{90}$Zr, ${}^{144}$Sm, and ${}^{208}$Pb. However, on 
closer examination these models appear to overestimate the GMR energy in all of the Tin isotopes. 
Indeed, whereas the theoretical spread among these model (not including NL3) amounts to at most
1\%, their predictions are incompatible with experiment. Thus, the question of {\sl ``why is Tin so 
soft?''} has been raised to the forefront of nuclear 
structure\,\cite{Piekarewicz:2008nh,Li:2007bp,Li:2010kfa,Piekarewicz:2007us,Sagawa:2007sp,Avdeenkov:2008bi,Piekarewicz:2009gb,Cao:2012dt}.
To underscore the challenge facing nuclear structure we display in Fig.\,\ref{Fig5} the distribution 
of isoscalar monopole strength in all stable, neutron-even Tin isotopes measured at 
RCNP\,\cite{Li:2007bp,Li:2010kfa}. Also shown in the figure are relativistic RPA
predictions obtained with the ``stiff'' NL3 and ``soft'' FSUGold interactions. Note that for clarity
the theoretical peak energy has been normalized to the experimental data. The RCNP experiment 
aimed to probe the incompressibility of neutron-rich matter by measuring the GMR along 
a chain of isotopes with a neutron excess in the range $\alpha\!=\!0.11$-$0.19$. Clearly discernible 
in the figure is the larger width of the experimental distribution relative to the theoretical predictions. 
This is expected, as RPA calculations are unable to account for the complicated spreading width. 
What is not expected is the systematic hardening of the 
theoretical predictions---especially given that FSUGold reproduces the centroid energies of 
${}^{90}$Zr, ${}^{144}$Sm, and ${}^{208}$Pb.
\begin{figure}[ht]
 \begin{center}
  \includegraphics[width=0.75\linewidth,angle=0]{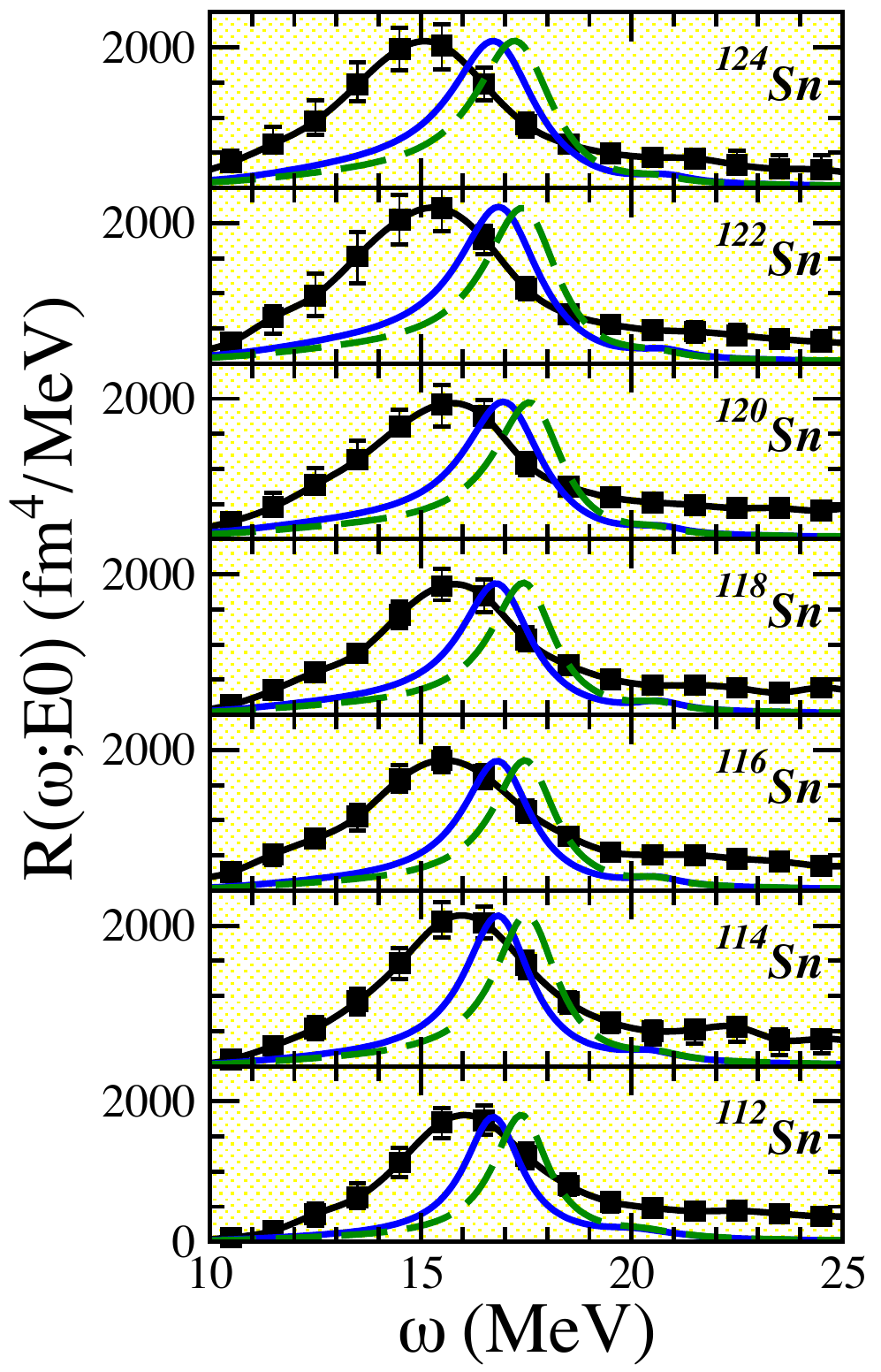}
  \caption{(color online) Comparison between the distribution of isoscalar monopole strength 
  in all stable, neutron-even Tin isotopes measured at RCNP\,\,\cite{Li:2007bp,Li:2010kfa} 
  against the theoretical predictions of NL3 (green dashed line) and FSUGold (blue solid line).} 
 \label{Fig5}
 \end{center} 
\end{figure}
To further elucidate this problem we display in Fig.\,\ref{Fig6} centroid energies obtained by 
integrating the distribution of monopole strength over the $\omega\!=\!10\!-\!20$\,MeV interval.
Two points are worth emphasizing: (a) the value of the centroid energy in ${}^{112}$Sn and (b)
the softening of the mode as a function of $A$. Given that the neutron excess in ${}^{112}$Sn
is small ($\alpha^{2}\!\approx\!0.01$), the value of its centroid energy is mostly sensitive to 
$K_{0}$ with only a 1\% correction from $K_{\tau}$. This is clearly reflected in the predictions
of NL3 (with $K_{0}\!=\!271.5$\,MeV) and FSUGold (with $K_{0}\!=\!230.0$\,MeV). In particular,
the FSUGold prediction is within 1\% of the (upper limit of the) experimental value. However, 
the experiment suggests a very rapid softening of the mode with mass number that is not 
reproduced by either NL3 or FSUGold. Note that the falloff with $A$ is controlled---at least in 
infinite nuclear matter---by $K_{\tau}$ so the softening predicted by NL3 is indeed faster than 
that of FSUGold, but clearly nowhere as fast as required by the experiment. Hence, the 1\% 
discrepancy ${}^{112}$Sn between FSUGold and experiment, increases by a factor of 4---or
to 0.7\,MeV which is significantly larger than the 0.1\,MeV experimental error. Also shown in 
Fig.\,\ref{Fig6} are the predictions from a ``Hybrid'' model that was created with the sole purpose 
of reproducing the RCNP data\,\cite{Piekarewicz:2008nh}. In particular, the Hybrid model has
the same soft incompressibility coefficient as FSUGold ($K_{0}\!=\!230$\,MeV) but a 
considerable stiffer symmetry energy ($K_{\tau}\!=\!-\!565$\,MeV). Note that these bulk
parameters agree with the recommended values of $K_{0}\!=\!240\!\pm\!10$\,MeV and 
$K_{\tau}\!=\!-550\!\pm\!100$\,MeV given in Ref.\,\cite{Li:2007bp}. Unquestionably, the
Hybrid model produces a significant improvement in the description of the data. Indeed, 
accounting for the experimental errors, the theoretical predictions fall within $0.1$~MeV 
of the experimental data along the full isotopic chain. However, the falloff with $A$ is still 
not as rapid as required by experiment. Yet, the Hybrid model suffers from an ever more 
serious problem: it underestimates the GMR centroid energy in ${}^{208}$Pb by almost 
1\,MeV; see also Ref.\,\cite{Avdeenkov:2008bi}. Given that the solution to the problem 
of the softness of Tin should not come at the expense of sacrificing the overall quality
of the model, we must conclude that the Hybrid model is unrealistic. 
\begin{figure}[ht]
 \begin{center}
  \includegraphics[width=0.9\linewidth,angle=0]{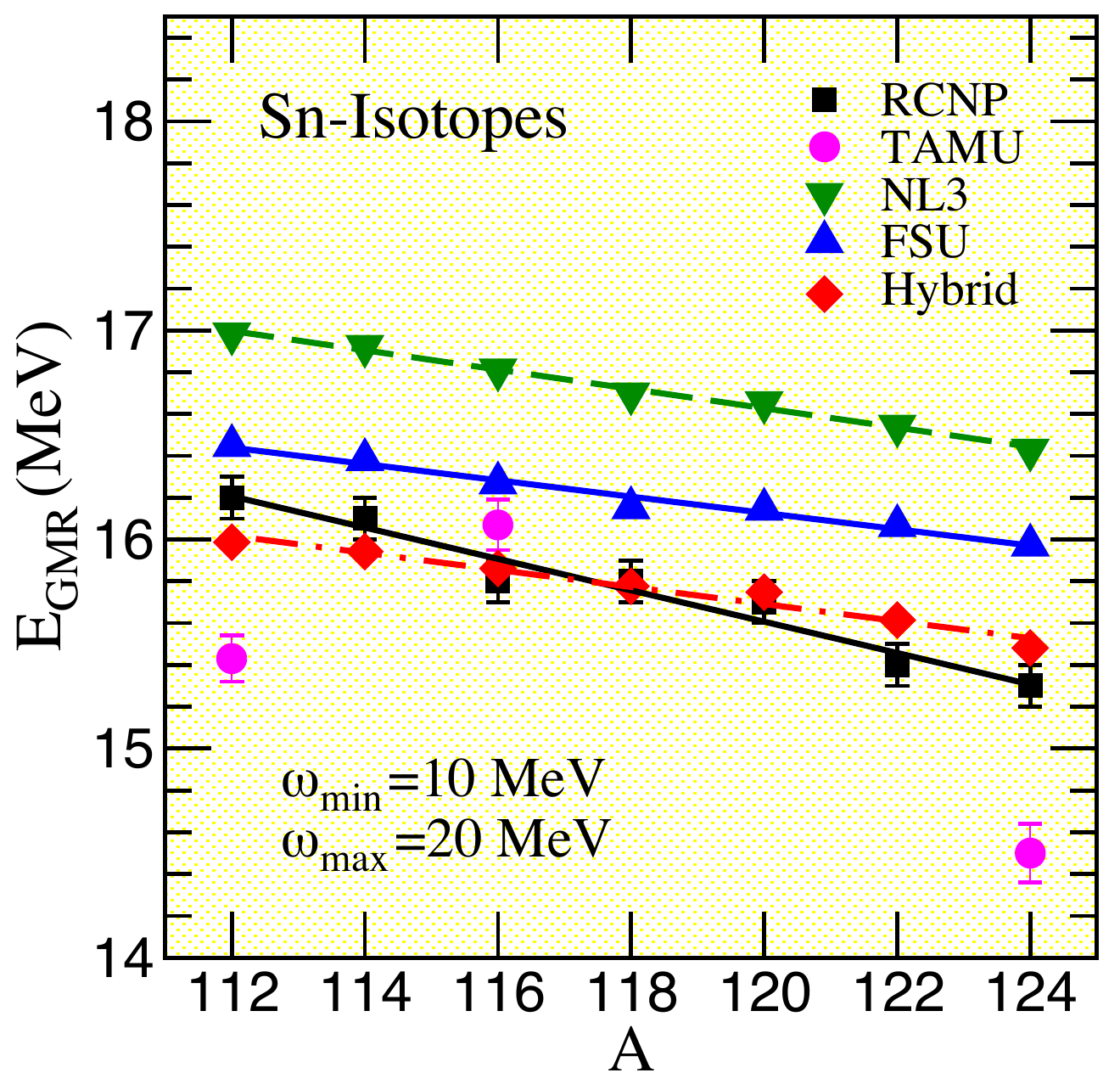}
    \caption{(color online) Comparison between the GMR centroid energies in all stable, 
    neutron-even Tin isotopes measured at RCNP\,\,\cite{Li:2007bp,Li:2010kfa} against 
    the theoretical predictions of NL3 (green down-triangles), FSUGold (blue up-triangles),
    and the Hybrid model (red diamonds) introduced in Ref.\,\cite{Piekarewicz:2008nh}.
    Also shown (magenta circles) are measurements from the Texas A\&M 
    group\,\cite{Youngblood:1999,Lui:2004wm} for ${}^{112,116,124}$Sn.} 
 \label{Fig6}
 \end{center} 
\end{figure}
So after more than 5 years since the publication of the experimental data, the answer to the 
question of ``why is Tin so soft?'' continues to elude us. Note that the softness of Tin has 
been recently confirmed in the nearby isotopic chain in Cadmium\,\cite{Patel:2012zd}. It
has been proposed that the failure to reproduce the GMR energies in Sn, and now in Cd, 
may be due to missing physics unrelated to the bulk incompressibility of neutron-rich
matter. One suggestion that has received considerable attention involves the superfluid 
character of the Tin 
isotopes\,\cite{Li:2008hx,Khan:2009xq,Khan:2009ih,Khan:2010mv,Vesely:2012dw}. 
Although the conclusions have been mixed and seem to depend strongly on the character 
of the pairing force, it is safe to assume that pairing correlations do not provide a robust
softening mechanism. Regrettably then, we are forced to conclude that the softness 
of both Tin and Cadmium remains an important open problem.

\subsection{Isovector Giant Dipole Resonance}
\label{res:ivgdr}

\begin{figure}[ht]
 \begin{center}
  \includegraphics[width=0.9\linewidth,angle=0]{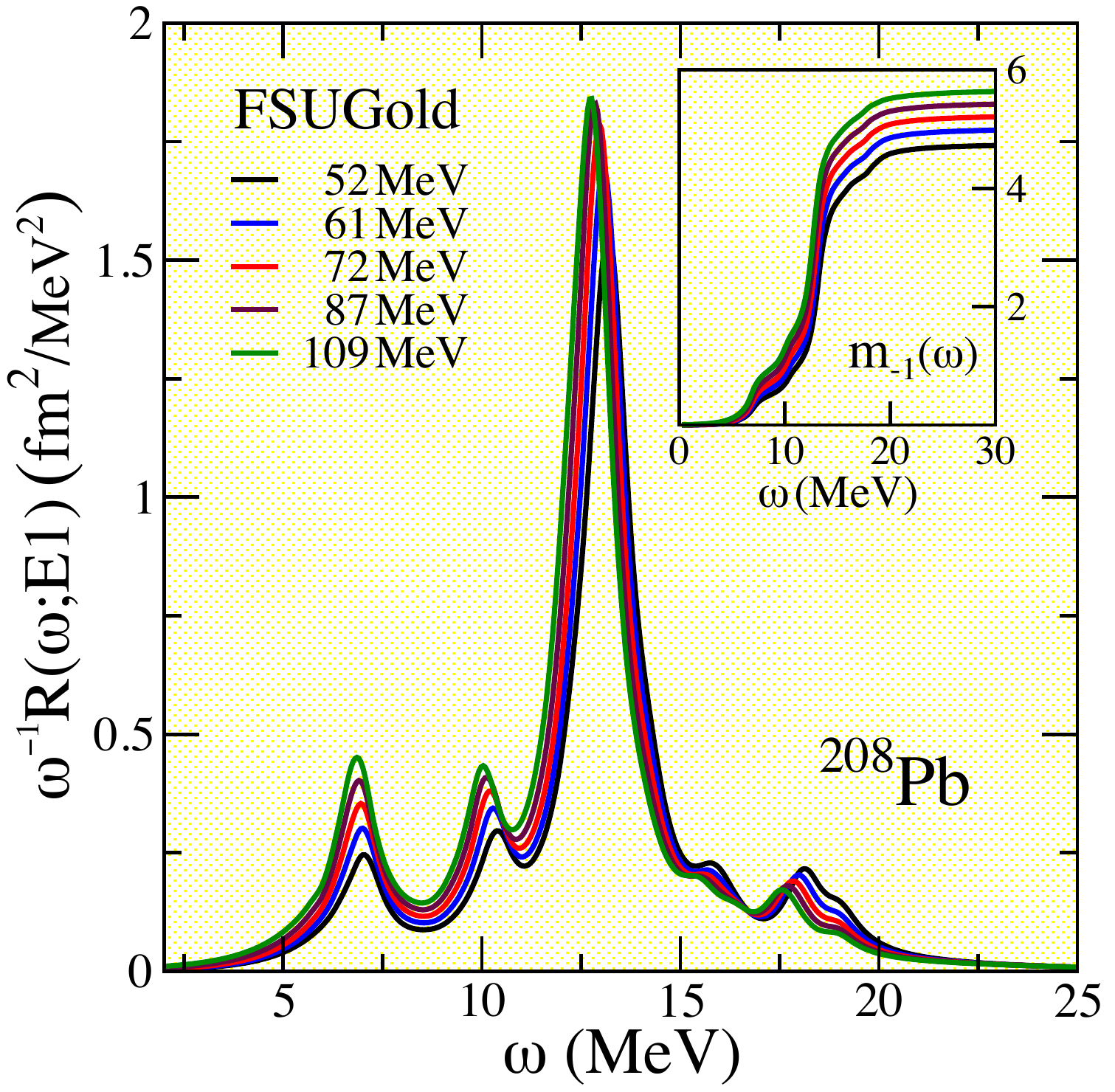}
  \caption{(color online) Distribution of inverse energy weighted isovector dipole
  strength for ${}^{208}$Pb as predicted by the FSUGold family of effective interactions. 
  The labels indicate the slope of the symmetry energy $L$ predicted by each model
  and the inset shows the running sum $m_{-1}(\omega)$ as defined in Eq.\,(\ref{Rsum}).}  
  \label{Fig7}
 \end{center} 
\end{figure}
Whereas probing to the symmetry energy via the GMR is hindered by the relative small 
neutron-proton asymmetry of stable nuclei, the isovector dipole mode provides a direct 
access to its density dependence. Indeed, the GDR is commonly perceived as an out
of phase oscillation of neutrons against proton with the symmetry energy acting as the 
restoring force. In particular, models with a stiff symmetry energy ({\sl i.e.,} models with
an energy that increases rapidly with increasing density) predict small values for the symmetry 
energy at the sub-saturation densities of relevance to the excitation of this mode. As a result, 
models with a stiff symmetry energy, and thus large values of $L$, predict a distribution 
of dipole strength that is both enhanced and softened relative to their softer counterparts. 
However, although the distribution of dipole strength is sensitive to the density dependence 
of the symmetry energy, the energy weighted sum is not. This is because the $m_{1}$ 
moment of the distribution is ``protected'' by the classical model independent EWSR 
[see Eq.\,(\ref{EWSR})]. A far better isovector indicator is the electric dipole 
polarizability which is directly proportional to the inverse 
EWSR\,\cite{Reinhard:2010wz,Piekarewicz:2010fa,Piekarewicz:2012pp,Roca-Maza:2013mla}.
The dipole polarizability is particularly attractive because here the softening and 
enhancement are intensified as one weighs the response with $\omega^{-1}$. 
This point is nicely illustrated in Fig.\,\ref{Fig7} which 
displays the inverse energy weighted distribution of isovector dipole strength in 
${}^{208}$Pb for the ``family" of FSUGold interactions. The virtue of introducing 
such family of interactions is that one can assess systematic changes in a given 
observable as a single bulk parameter of the EOS is modified. In the particular case 
of the FSU family of interactions, one modifies the density dependence of the 
symmetry energy---essentially the value of $L$---while leaving the isoscalar sector 
intact\,\cite{Horowitz:2001ya}. Indeed, all models displayed in Fig.\,\ref{Fig7} share 
the same value of $\rhoz$, $\epsz$, $K_{0}$, and $Q_{0}$. Note that the models 
displayed in the figure are labeled according to their value of $L$. As mentioned
earlier, the stiffer the symmetry energy the larger and softer the response. This
trend becomes evident in the inset to the figure which displays the running sum 
$m_{-1}(\omega)$ defined as
\begin{equation}    
  m_{-1}(\omega) \equiv \int_{0}^{\omega}\!z^{-1}R(z;E1)\, dz \;.
 \label{Rsum}
\end{equation}
Although there is already some separation at lower excitation energies, by the 
time that one has integrated over the region of the GDR the model dependence 
becomes apparent. Hence a powerful correlation emerges: the larger the value 
of $L$, the larger the dipole polarizability in ${}^{208}$Pb. Moreover, this correlation 
may be framed in the form of a powerful ``data-to-data'' relation: the larger the 
neutron-skin thickness of ${}^{208}$Pb the larger its dipole polarizability. 

\begin{figure}[t]
 \begin{center}
  \includegraphics[height=7.50cm,width=7.00cm]{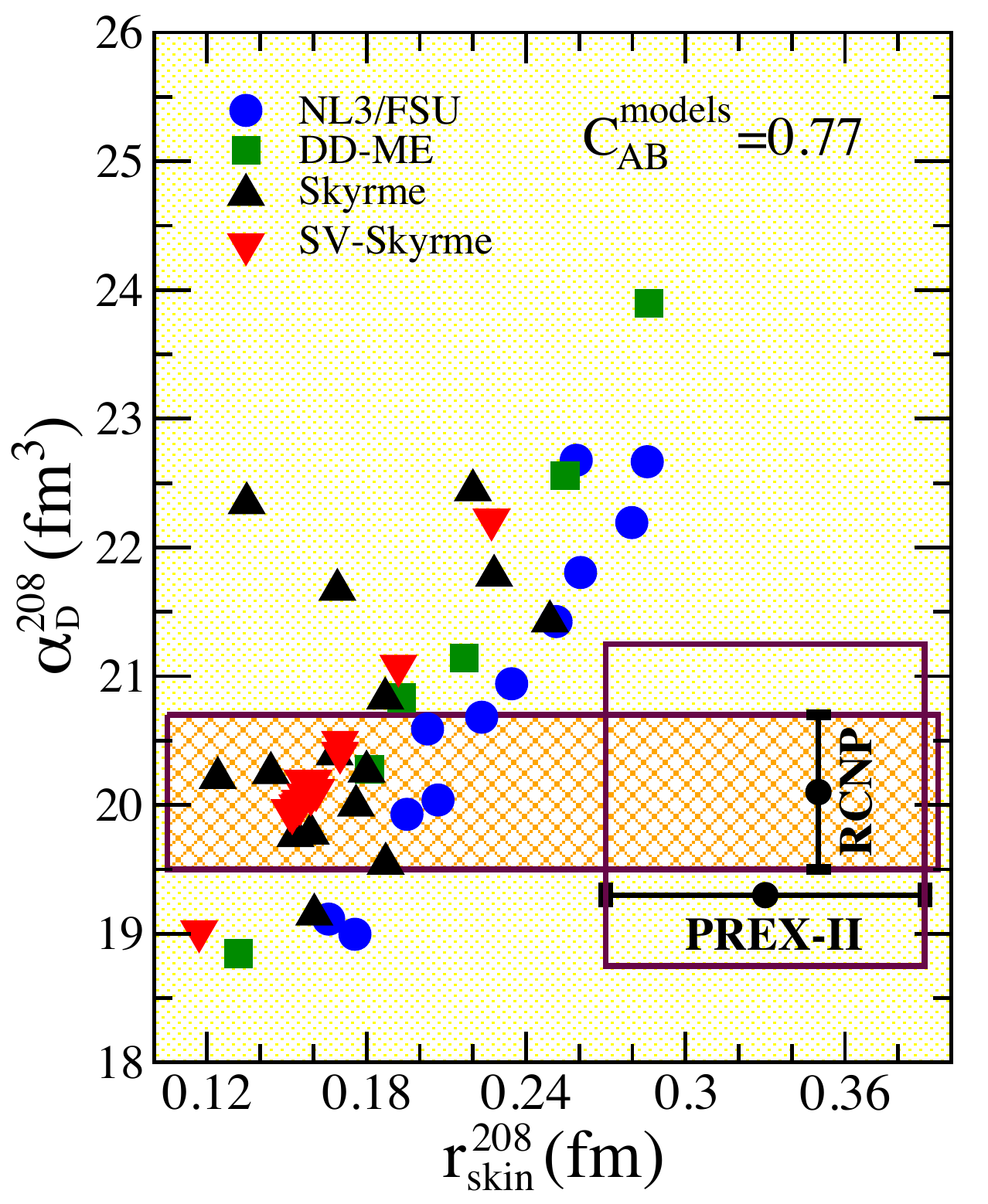}
  \caption{(Color online) Predictions from 48 nuclear EDFs for the electric 
  dipole polarizability and neutron-skin thickness of ${}^{208}$Pb as
  discussed in Ref.\,\cite{Piekarewicz:2012pp}. Constrains from both
  RCNP\,\cite{Tamii:2011pv,Poltoratska:2012nf} and PREX\,\cite{Abrahamyan:2012gp} 
  (the latter assuming a projected 0.06\,fm error) have been incorporated into the plot.}
 \label{Fig8}
 \end{center} 
\end{figure}
Such a data-to-data relation between the neutron-skin thickness and electric dipole 
polarizability in ${}^{208}$Pb is displayed in Fig.\,\ref{Fig8} for a large set of 48 relativistic 
and non-relativistic energy density functionals\,\cite{Piekarewicz:2012pp}. Even though a
clear linear trend between $\alphad^{208}$ and $\rskin^{208}$ is observed, one also
notices a significant scatter in the predictions of these models. However, note that 
each set of systematically varied models (all models with the exception of the standard 
Skyrme interactions depicted with the black triangles) display an almost perfect correlation. An
overall correlation coefficient of 0.77 is found when the predictions from all 48 EDFs 
are included. Note that by imposing the recent experimental constraint from 
$\alphad^{208}$\,\cite{Tamii:2011pv,Poltoratska:2012nf} several models, especially 
those with a very stiff symmetry energy, may already be ruled out. Remarkably, if the 
updated PREX experiment (PREX-II)---with a projected uncertainty of 0.06\,fm---finds 
that its central value of $\rskin^{208}\!=\!0.33$\,fm remains intact, then all 48 models 
displayed in the figure will be ruled out!

Although the correlation between $\alphad^{208}$ and $\rskin^{208}$ displayed in Fig.\,\ref{Fig8}
is evident, insights from the droplet model [Eq.\,(\ref{dpdm})] suggest that a far better 
correlation involves the product of $\alphad^{208}$ times the symmetry energy at saturation 
density $J$ predicted by each model\,\cite{Roca-Maza:2013mla}. Thus, we display in 
Fig.\,\ref{Fig9} the product $\alphad^{208}\!J$ as a function of both $\rskin^{208}$ (in the 
left-hand panel) and $L$ (in the right-hand panel). Remarkably, the large spread in the model 
predictions seen in Fig.\,\ref{Fig8} has been practically eliminated. Indeed, the
figure validates (with correlation coefficients of $0.97$ and $0.96$) that the product 
$\alphad^{208}\,\!J$ is far better correlated to both $\rskin^{208}$ and $L$ than the polarizability 
alone. Note that the set of density functionals used in Fig.\,\ref{Fig9} now includes two additional 
EDFs, namely, TAMUC-FSU\,\cite{Fattoyev:2013yaa} and SAMi\,\cite{RocaMaza:2012sj}, with 
the former having a stiff and the latter a soft symmetry energy. Between them, they span a large range 
in neutron-skin thickness: $\rskin^{208}\!\approx\!(0.12$-$0.33$)\,fm or equivalently 
$L\!\approx\!(30$-$135$)\,MeV. Still, the predictions from both of these models fall comfortably 
within the confidence bands. We expect that during the next few years high-precision 
measurements of the electric dipole polarizability on a variety of nuclei will provide stringent 
constraints on the density dependence of the symmetry energy.
\begin{figure*}[ht]
 \begin{center}
  \includegraphics[height=6.75cm,width=8.25cm]{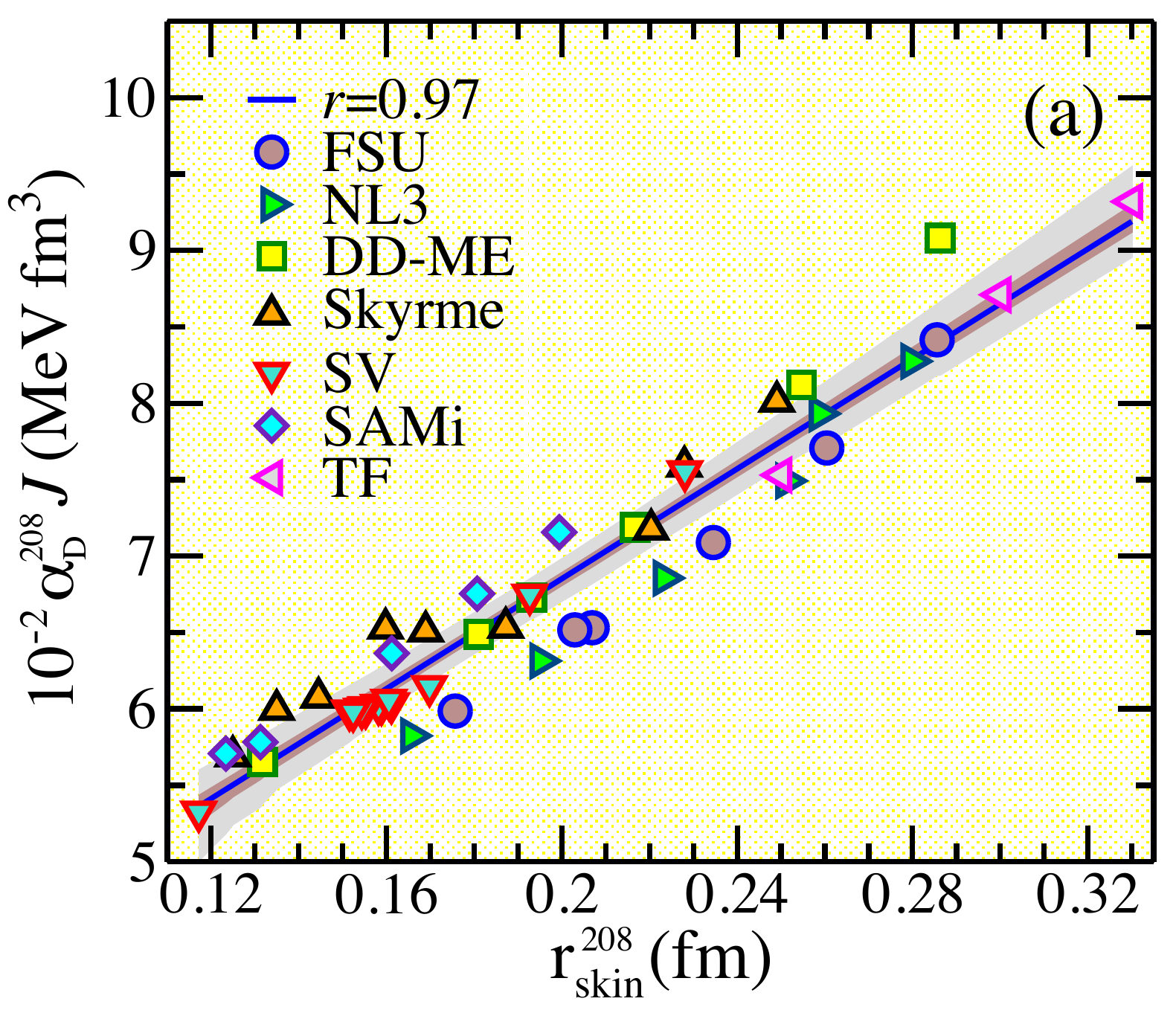}
  \hspace{0.5cm}
  \includegraphics[height=6.75cm,width=8.25cm]{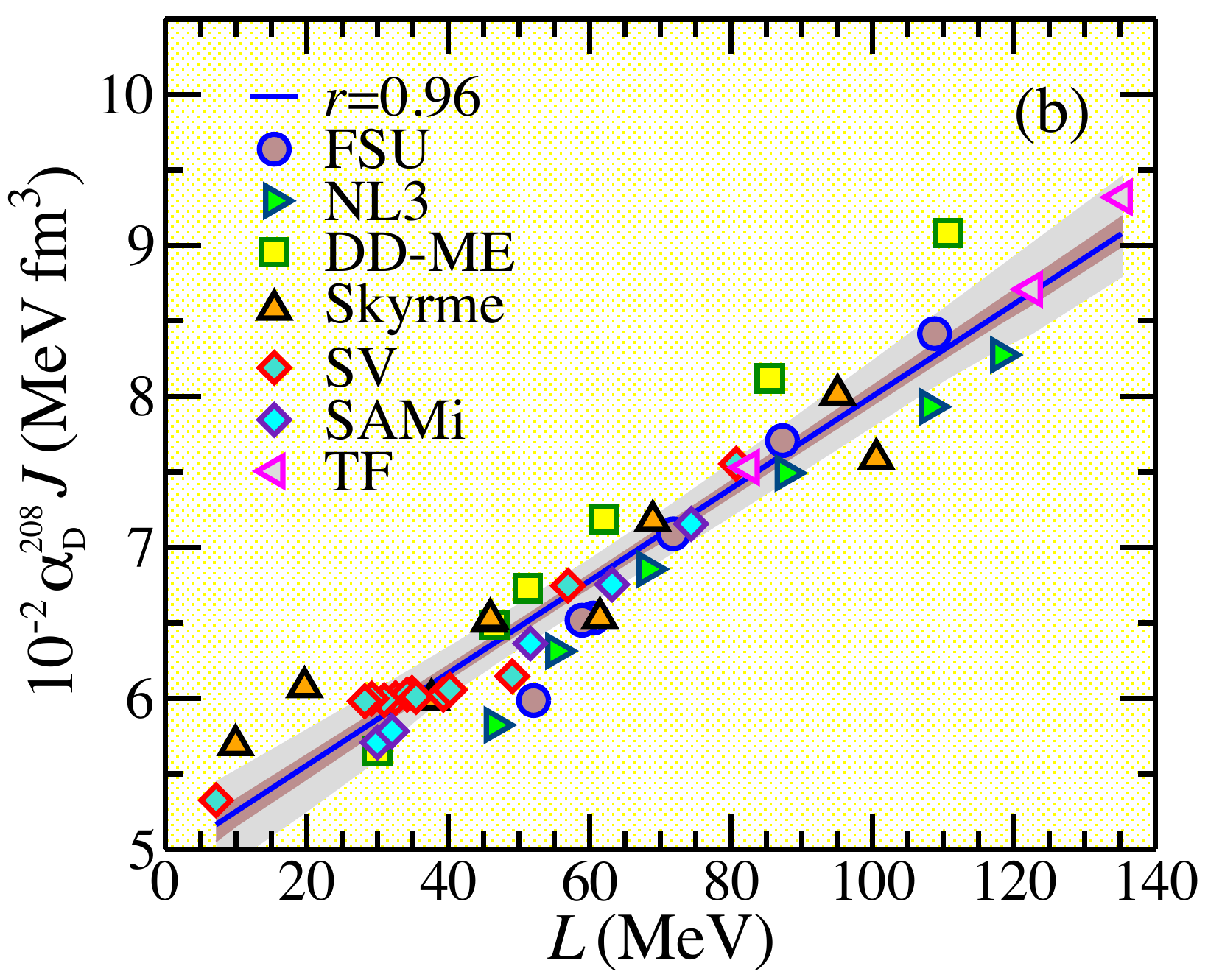}
  \caption{(Color online) (a) Predictions from a large number of EDFs for the electric 
  dipole polarizability of ${}^{208}$Pb times the symmetry energy at saturation density
  as a function of the neutron-skin thickness of ${}^{208}$Pb. (b) Same as in (a) but
  now as a function of the slope of the symmetry energy $L$. The two shaded regions 
  represent the 99.9\% and 70\% confidence bands, respectively. These results were
  first reported in Ref.\,\cite{Roca-Maza:2013mla}.} 
 \label{Fig9}
 \end{center} 
\end{figure*}

\subsubsection{Pygmy Dipole Resonance}
\label{pygmy}

We conclude this section by discussing the interesting and often controversial topic of low energy 
dipole strength, the so-called Pygmy Dipole Resonance (PDR). Proposals on the possible 
existence of a new type of dipole oscillation in which the neutron skin vibrates against the 
isospin symmetric core date back to the early 1990's\,\cite{Suzuki:1990,VanIsacker:1992}. 
Since then, state-of-the-art RPA calculations have been developed to predict the distribution
of of low-energy dipole strength and associated transition densities for a variety of neutron-rich 
nuclei\,\cite{Hamamoto:1996,Hamamoto:1998,Vretenar:2000yy,Vretenar:2001hs,Paar:2002gz,Tsoneva:2003gv,Sarchi:2004,Paar:2004gr}.
Given that stable heavy nuclei are neutron rich, one expects the emergence of low energy 
dipole strength as the nucleus develops a neutron-rich skin. Thus, it was suggested that the 
PDR may be used as a constraint on the neutron skin of heavy 
nuclei\,\cite{Tsoneva:2003gv} as well as on the density dependence of the 
symmetry energy, and ultimately on the properties of neutron stars\,\cite{Piekarewicz:2006ip}. In particular, 
the fraction of the energy weighted sum rule exhausted by the PDR was shown to be sensitive 
to the neutron-skin thickness of heavy nuclei~\cite{Carbone:2010az,Tsoneva:2003gv,Piekarewicz:2006ip,Tsoneva:2007fk,Klimkiewicz:2007zz}.  
Pioneering experiments on unstable neutron-rich isotopes in Sn, Sb, and Ni seem
to support this assertion\,\cite{Klimkiewicz:2007zz,Adrich:2005,Wieland:2009,Rossi:2012ew}.
For a recent review on the Pygmy Dipole Resonance see Ref.\,\cite{Savran:2013bha}.

\begin{figure*}[ht]
 \begin{center}
  \includegraphics[height=7.25cm,width=8.25cm]{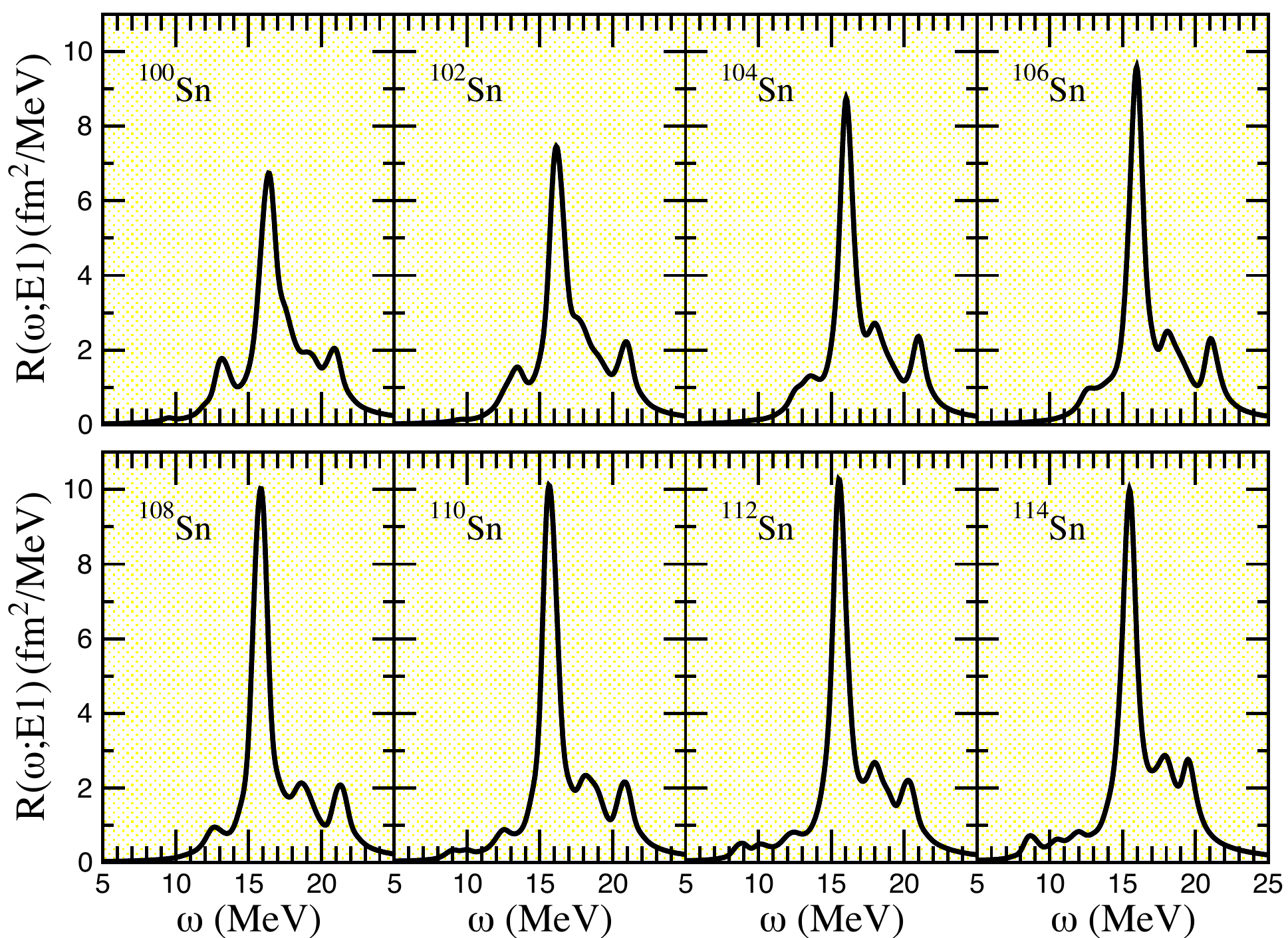}
  \hspace{0.0cm}
  \includegraphics[height=7.24cm,width=8.25cm]{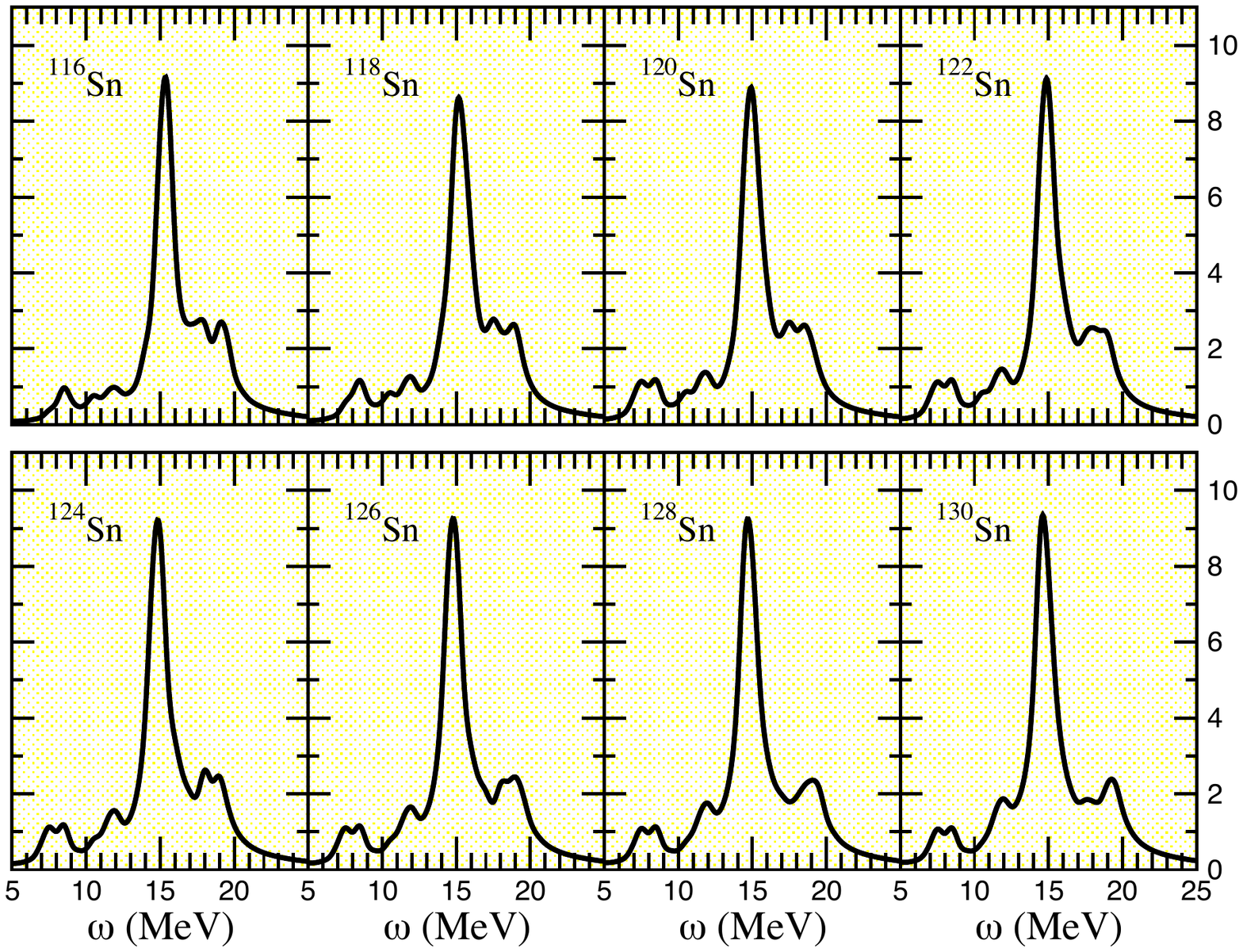}
  \caption{(Color online) Distribution of isovector dipole strength for all neutron-even tin 
  isotopes from ${}^{116}$Sn to ${}^{130}$Sn. Relativistic RPA calculations were carried
  out with the FSUGold effective interaction\,\cite{Todd-Rutel:2005fa}.}
 \label{Fig10}
 \end{center} 
\end{figure*}

To observe the emergence of pygmy strength as the nucleus develops a neutron-rich skin, 
we display in Fig.\,\ref{Fig10} the distribution of isovector dipole strength $R(\omega;E1)$ 
for all neutron-even Sn isotopes from ${}^{100}$Sn to ${}^{130}$Sn. The results shown in
the figure are obtained from RPA calculations using the FSUGold effective 
interaction\,\cite{Todd-Rutel:2005fa}. The large collective structure in the $\omega\!\gtrsim\!15$\,MeV 
corresponds to the giant dipole resonance. The collective character of the GDR is associated 
with the coherent out of phase oscillation of neutrons against protons and is observed in all 
isotopes regardless of neutron excess. Moreover, as is characteristic of such a giant resonance,
it exhausts a large fraction of the EWSR. But clearly not all! The emergence of low-energy
dipole strength with increasing neutron excess is clearly discernible in the figure. Indeed, 
the progressive addition of neutrons results in a well developed, albeit 
small, low-energy resonance. Note that the PDR is barely visible below ${}^{108}$Sn, increases 
gradually beyond ${}^{110}$Sn, and finally saturates at ${}^{120}$Sn. This saturation effect is 
attributed to the filling of the $1h^{11/2}$ neutron orbital. Whereas the 12 neutrons filling 
the $1h^{11/2}$ orbital contribute to a systematic increase in the size of the neutron skin, 
high-angular momentum orbitals play a minor role in generating low-energy dipole 
strength\,\cite{Piekarewicz:2006ip}.
\begin{figure}[ht]
 \begin{center}
  \includegraphics[width=0.9\linewidth,angle=0]{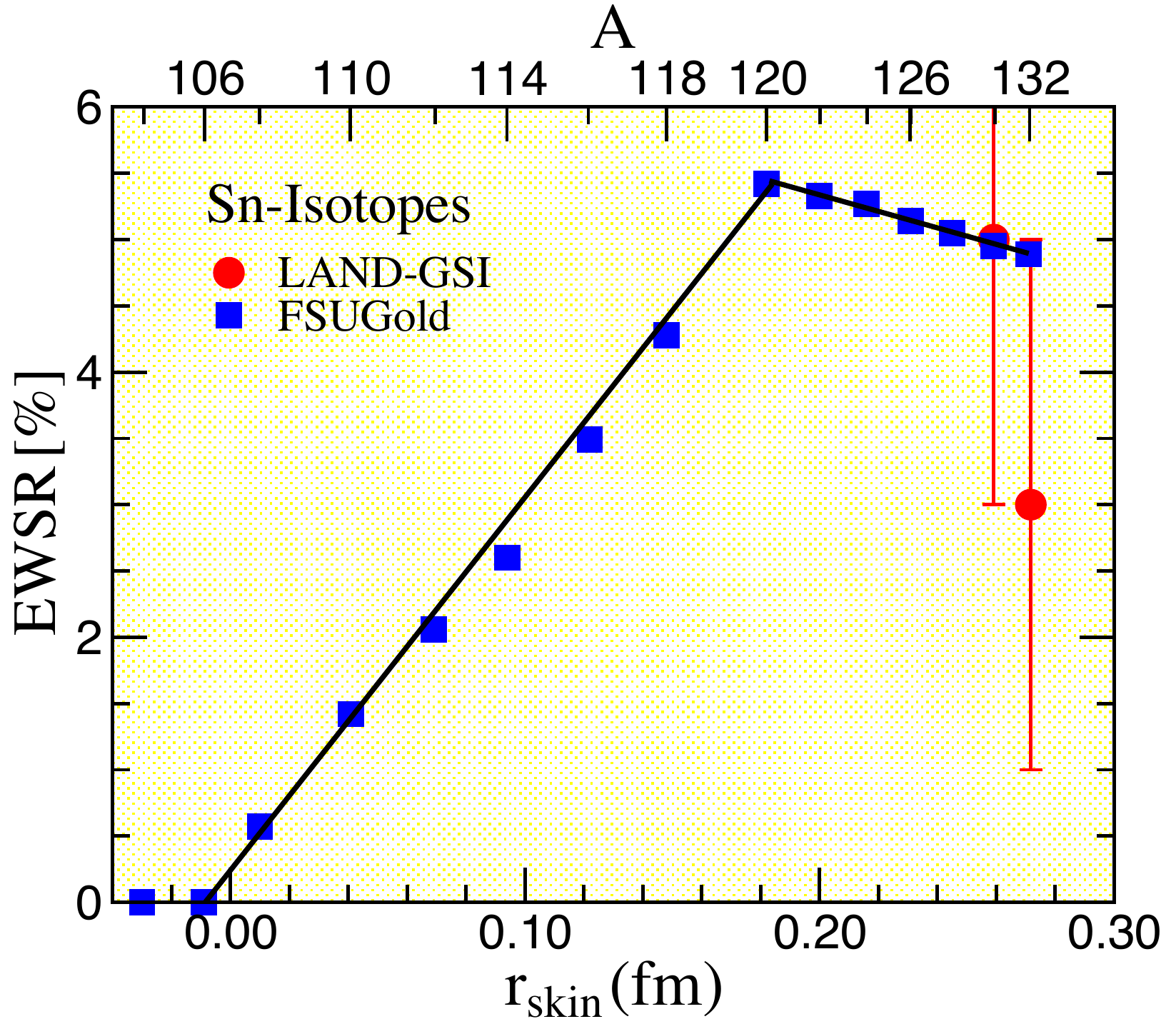}
  \caption{(Color online) Fraction of the energy weighted sum rule contained in the PDR 
  relative to that in the GDR for the isotopic chain in Sn. The experimental data is from 
  Ref.\,\cite{Adrich:2005}.}
 \label{Fig11}
 \end{center} 
\end{figure}

The fraction of the EWSR contained in the PDR relative to that of the GDR is displayed in 
Fig.\,\ref{Fig11} as a function of neutron-skin thickness. Note that below ${}^{108}$Sn, where
there is no visible low-energy strength, the neutron-skin thickness of the various Sn-isotopes
is negative. The experimental data is from the pioneering measurement on the unstable 
${}^{130,132}$Sn isotopes carried out by the LAND-FRS collaboration at 
GSI\,\cite{Adrich:2005}. The separation between low- and high-energy strength was 
chosen at $\omega\!=\!10$\,MeV. The linear correlation between $r_{\rm skin}$ and the fraction 
of dipole strength exhausted by the PDR justifies the assertion that the neutron excess is
indeed responsible for the emergence of low-energy strength as the nucleus becomes
progressively more neutron rich. Moreover, the saturation of this effect due to the failure 
of the $1h^{11/2}$ neutron orbital to generate low-energy strength is also clearly visible.
Unfortunately, at present the experimental error bars are too large to place significant 
constraints on the density dependence of the symmetry energy. However, by combining 
information on the the fraction of the EWSR exhausted by the PDR in both 
${}^{132}$Sn\,\cite{Adrich:2005} and ${}^{68}$Ni\,\cite{Wieland:2009}, Carbone and 
collaborators were able to constrain the slope of the symmetry energy to the range
of $L\!=\!64.8\pm15.7$\,MeV\,\cite{Carbone:2010az}.

In spite of this success, we believe that the reliance on the energy weighted sum rule
suffers from two serious limitations. First, the classical EWSR is model independent so the
sensitivity of the fraction exhausted by either the PDR or GDR to the symmetry energy 
must be modest. Second, in generating the EWSR one weighs the distribution of strength 
with the energy, thereby diminishing the impact of the PDR even further. In contrast, the
inverse energy weighted sum rule does not suffer from any of these problems. First, as 
shown in Fig.\,\ref{Fig7}, the inverse EWSR is sensitive to the density dependence of the 
symmetry energy. Second, the low-energy strength is now enhanced by weighting the 
distribution of strength with $\omega^{-1}$. A nucleus with a significant amount of low-energy 
dipole strength is the exotic, neutron-rich ${}^{68}$Ni isotope\,\cite{Wieland:2009,Rossi:2012ew}. 
\begin{figure*}[ht]
 \begin{center}
  \includegraphics[height=7.5cm,width=7.25cm]{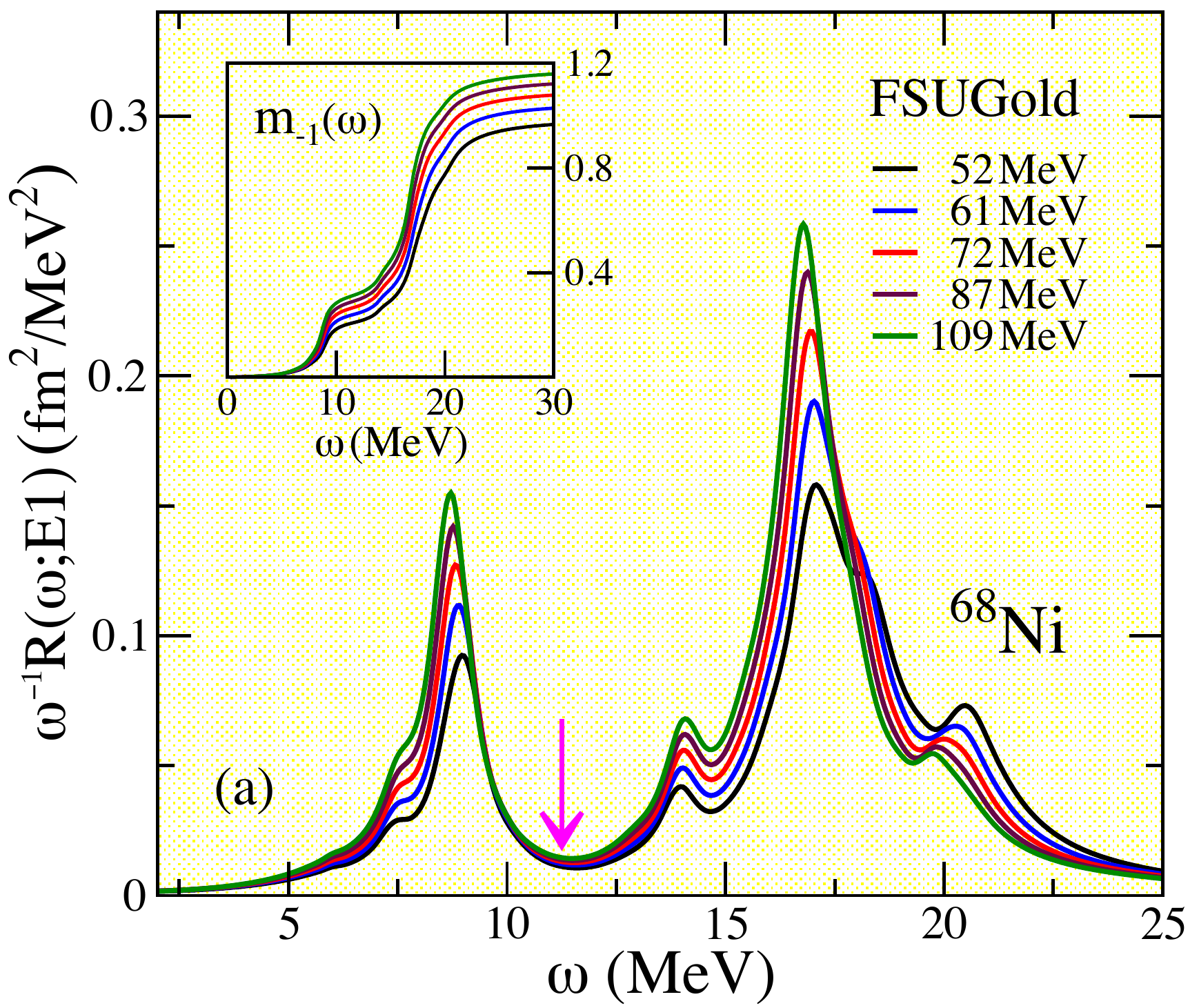}
  \hspace{0.0cm}
  \includegraphics[height=7.5cm,width=6.50cm]{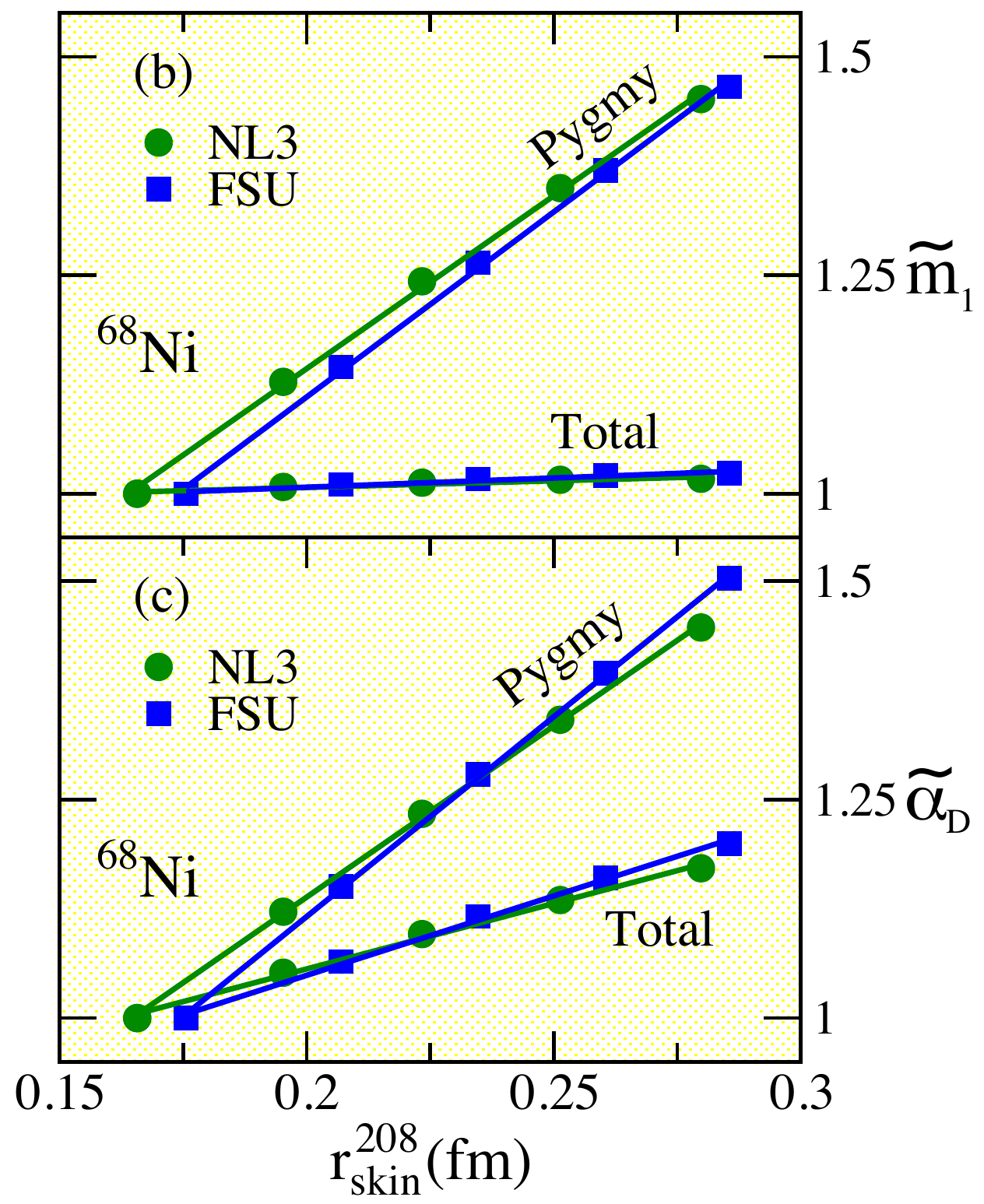}
  \caption{(color online) (a) Distribution of inverse energy weighted isovector dipole
  strength in ${}^{68}$Ni as predicted by the FSUGold family of effective interactions. 
  The labels indicate the slope of the symmetry energy $L$ predicted by each model
  and the inset shows the running sum $m_{-1}(\omega)$ as defined in Eq.\,(\ref{Rsum}).
  (b) Fractional change in the energy weighted sum rule of ${}^{68}$Ni as a function of the 
  neutron-skin thickness of ${}^{208}$Pb as predicted by both NL3 and FSUGold. (c) same
  as (b) but now for the electric dipole polarizability.}  
 \label{Fig12}
 \end{center} 
\end{figure*}
The inverse energy weighted distribution of isovector dipole strength in ${}^{68}$Ni is displayed on
the left-hand panel of Fig.\,\ref{Fig12} for the family of FSUGold interactions. For reference, the
fraction of the EWSR exhausted by the PDR in ${}^{68}$Ni ranges from about 5\% for the softest
model to about 8\% for the stiffest. In contrast, the fraction of the inverse EWSR contained in the
PDR is significantly larger and varies in the range 20-25\%; note that the corresponding fraction in 
${}^{208}$Pb---although still sizable---amounts to only 10\% [see  Fig.\,\ref{Fig7}]. Moreover, the 
running sum $m_{-1}(\omega)$ depicted in the inset shows that there is significant model sensitivity
to the amount of strength contained in both the PDR and the GDR. Although somehow arbitrary, the 
arrow located in the ``dip region'' is used to define the energy separating the low-energy from the
high-energy region. To highlight these features we display in Figs.\,\ref{Fig12}b and\,\ref{Fig12}c
NL3 and FSUGold predictions for the fractional changes in the total and Pygmy contributions to 
both $m_{1}$ and $\alphad$ as a function of the neutron-skin thickness of ${}^{208}$Pb. Note that 
for ease of view we have plotted the fractional changes, {\sl i.e,} we have normalized the lowest
point to one, and have used a {\sl ``tilde''} to denote these fractional changes. As anticipated, given that 
$m_{1}$ satisfies a classical sum rule, the sensitivity of the total EWSR to the density dependence 
of the symmetry energy is very weak; the slight deviation from one is due to the model dependence 
of the TRK enhancement factor. Conversely, the PDR displays a large sensitivity, albeit with a 
small signal, and could in principle be used to place constraints on the neutron-skin thickness of ${}^{208}$Pb. 
Indeed, Carbone {\sl et al.} have used measurements of the fraction of the EWSR exhausted by the 
PDR in both ${}^{132}$Sn\,\cite{Adrich:2005} and ${}^{68}$Ni\,\cite{Wieland:2009} to extract 
a value of $r_{\rm skin}^{208}\!=\!0.194\pm\!0.024$\,fm\,\cite{Carbone:2010az}. However, 
the dipole polarizability enjoys several advantages over the EWSR. First, as far as the PDR 
is concerned, the sensitivity to $r_{\rm skin}^{208}$ is as large for $\alphad$ as it is for $m_{1}$. 
Second, the total dipole polarizability is not protected by a classical sum rule. Thus, unlike the EWSR, 
the sensitivity of the total dipole polarizability to $r_{\rm skin}^{208}$ is significant. Finally---and 
most importantly---the total dipole polarizability is immune to the arbitrary separation of low-energy 
from high-energy strength. Indeed, whereas Fig.\,\ref{Fig12}a shows a clear separation, experimentally 
the GDR tends to be quite broad, so deciding how much strength is contained in the PDR and how much 
in the tail of the GDR is likely to be ill defined\,\cite{Savran:2013bha}. Thus, we conclude that the 
(total) electric dipole polarizability is a robust isovector indicator that provides critical insights
into the density dependence of the symmetry energy. Nevertheless, the PDR is enormously interesting 
and one should devote time and resources to investigate its character and its connection, if any, to the
neutron-skin thickness of heavy nuclei. Regardless, we believe that independent of the nature of the PDR, 
the emergence of low-energy dipole strength as nuclei develop a neutron-rich skin is an incontrovertible 
fact\,\cite{Piekarewicz:2010fa}.

\section{Conclusions and Outlook}
\label{conclusions}
In this review we have examined the critical role played by giant resonances in constraining the
density dependence of the symmetry energy. In particular, we focused on the isoscalar monopole 
and isovector dipole resonances. We made this choice in response to the enormous experimental 
effort devoted to these resonances and for the critical physical insights that have emerged as a 
result. 

Given that symmetric nuclear matter saturates, the incompressibility coefficient $K_{0}$ controls
the energetics of the small oscillations around the saturation density. The GMR---or nuclear breathing 
mode---is our window into the incompressibility of nuclear matter. However, the accurate determination 
of $K_{0}$ requires the formation of a strong collective peak that exhausts most of the 
energy weighted sum rule. As such, one must rely on heavy nuclei, which are necessarily neutron-rich.
This implies that experiments that measure the GMR in heavy nuclei (such as ${}^{208}$Pb) probe the 
incompressibility coefficient of neutron-rich matter, rather than that of symmetric matter. We
showed in Eq.\,(\ref{Kalpha}) that in terms of the neutron-proton asymmetry $\alpha$, the incompressibility
coefficient of asymmetric nuclear matter is given by $K({\alpha})\!=\!K_{0}\!+\!K_{\tau}\alpha^{2}$. The 
sensitivity of GMR energies to the density dependence of symmetry energy comes through its dependence 
on $K_{\tau}$. Indeed, $K_{\tau}$ is dominated by the slope of the symmetry energy $L$. Thus, it was 
believed that by measuring GMR centroid energies in ${}^{90}$Zr and ${}^{208}$Pb---nuclei with  
well-developed GMR peaks but values of $\alpha^{2}$ that differ by a factor of four---one could 
simultaneously constrain $K_{0}$ and $K_{\tau}$. Indeed, it was found that models with a relatively
small $K_{0}$  and a soft the symmetry energy could reproduce experimental
GMR energies in ${}^{90}$Zr, ${}^{144}$Sm, and ${}^{208}$Pb. However, after a short period of 
relative calm, an experiment at the RCNP facility in Japan has muddled the waters. The goal of 
the experiment was to measure the distribution of monopole strength in all even-A, stable nuclei along 
the isotopic chain in Tin. The great advantage of such an experiment is that one could monitor 
the evolution of GMR strength as the isotope becomes progressively more neutron rich. However,
the measurements revealed that self-consistent models that were successful in reproducing 
GMR energies in ${}^{90}$Zr, ${}^{144}$Sm, and ${}^{208}$Pb overestimate the energy in all 
Sn-isotopes. Moreover, the discrepancy between theory and experiment appears to grow 
with neutron excess. Finally, we note that the softness of Tin has been extended to the 
neighboring isotopic chain in Cadmium. Hence, we must conclude---in spite of considerable 
effort---that the question of {\sl ``why is Tin so soft?''} remains unanswered.

In the case of the isovector dipole resonance, the symmetry energy acts as the restoring force. 
Given that the excitation of this mode results in the spatial separation 
of two dilute quantum fluids, one neutron rich and the other one proton rich, the GDR probes the 
symmetry energy at sub-saturation densities. This suggests that the distribution of dipole strength
for models with a stiff symmetry energy is soften and enhanced relative to models having a 
soft symmetry energy. However, this sensitivity is eroded as one computes the energy weighted
sum, which is protected by a classical sum rule. Instead, an observable that is particularly sensitive
to the symmetry energy is the electric dipole polarizability (or inverse energy weighted sum) as
here the softening and the enhancement act in tandem. Indeed, a powerful data-to-data relation
may be established in this case: the thicker the neutron-skin thickness of ${}^{208}$Pb the larger 
the dipole polarizability. We note, however, that although not strictly a data-to-data relation, a far
stronger correlation emerges as one multiplies the dipole polarizability by the symmetry energy 
at saturation density [see Fig.\,\ref{Fig9}]. Finally, we addressed the intriguing role of the pygmy
dipole resonance in constraining the density dependence of the symmetry energy. Undeniably, 
we find a strong correlation between the emergence of low-energy dipole strength and the
development of a neutron-rich skin along the isotopic chain in Tin. What is unclear at this
juncture---but should be explored further---is whether one could cleanly separate the low
energy pygmy strength from the tail of the giant dipole resonance. Regardless of whether
this is possible, the total electric dipole polarizability was found to be a strong isovector indicator.

In summary, the distribution of isoscalar monopole strength has been found to be sensitive
to the density dependence of the symmetry energy. Unfortunately, this sensitivity is hindered 
by the relatively small neutron excess of the stable nuclei explored to date. In addition, we 
continue to puzzle after more than 5 years over the question of ``why is Tin so soft?'' Based
on these facts, we strongly encourage measurement of the distribution of monopole strength 
in exotic nuclei at next generation facilities. In particular, mapping centroid energies outside
the stable ${}^{112}$Sn to ${}^{124}$Sn region is sure to prove invaluable. Moreover, the
electric dipole polarizability $\alphad$ has been found to be sensitive to the density dependence 
of the symmetry energy. Moreover, the combination of $\alphad^{208}J$ was seen to be very
strongly correlated (with a correlation coefficient of 0.97) to the neutron-skin thickness of
$r_{\rm skin}^{208}$. Although photo-absorption experiments have been used for decades to 
probe the structure of the GDR, it is critical to delve into the low-energy region for a proper  
evaluation of $\alphad$; in ${}^{68}$Ni the PDR alone accounts for 25\% of the dipole
polarizability. Here too the systematic exploration of the isovector dipole strength along the 
chain of Sn-isotopes, for both stable and exotic nuclei, may prove priceless. In regards to 
theory, the need to provide meaningful uncertainties in theoretical predictions of physical 
observables is a theme that is gaining significant momentum among the scientific community. 
Indeed, the search for an accurately calibrated  microscopic theory that both predicts and 
provides well-quantified theoretical uncertainties is one of the founding pillars of modern 
nuclear energy density functionals\,\cite{UNEDF,Kortelainen:2010hv}. See the 
contribution to this volume by Nazarewicz, Reinhard, Satula, and Vretenar. The need to quantify 
model uncertainties in an area such as theoretical nuclear physics is particularly urgent as 
models that are fitted to experimental data are then used to extrapolate to the extremes of 
temperature, density, isospin asymmetry, and angular momentum. Inspired by some
critical insights and ideas developed in the context of non-relativistic Skyrme 
functionals\,\cite{Reinhard:2010wz}, a systematic statistical approach has been 
recently extended to the relativistic domain\,\cite{Fattoyev:2011ns,Fattoyev:2012rm}. 
An ambitious program aimed at calibrating future EDFs using ground-state properties
of finite nuclei, their collective response, and neutron-star properties---supplemented
by a proper covariance analysis---is well on its way\,\cite{Chen:2013tca}. A promising
first step in this direction has been taken recently\,\cite{Erler:2012qd}. 

We conclude our contribution with some thoughts on the role of other physical 
observables addressed in this volume in constraining the density dependence of the 
symmetry energy. Unquestionably, such a challenging task will require a coherent 
effort involving the theoretical, experimental, and observational communities. From 
a theoretical perspective, powerful theoretical constraints that have emerged from the 
nearly universal behavior of pure neutron matter at very low densities have provided a
stringent limit on the slope of the symmetry 
energy\,\cite{Hebeler:2010jx,Steiner:2011ft,Hebeler:2013nza}. See
the contributions to this volume by Gandolfi {\sl et al.} and Hebeler {\sl et al.} 
Experimentally, we have already discussed at length the merits of giant 
resonances, especially the dipole polarizability, in constraining the symmetry energy.
Together with the electric dipole polarizability, the neutron-skin thickness of heavy nuclei 
provide a powerful set of experimental constraints on the symmetry energy. The critical role
of electroweak measurements of neutron densities is discusses in this volume by
Horowitz, Kumar, and Michaels. Suffice it to say that elastic electron scattering 
is particularly advantageous as it provides a clean probe of neutron densities that is 
free from strong-interaction uncertainties. Finally, many contributions to this volume
have addressed the imprint of the symmetry energy on astrophysical observables.
Particularly promising is the accurate determinations of neutron-star radii---an observable
sensitive to the density dependence of the symmetry energy at intermediate and high
densities. Undoubtedly, enormous advances in land- and spaced-based observatories
have brought us closer to one of the holy grails of neutron-star physics: mapping
the mass-radius relation (see the contribution by Lattimer and Steiner to this
volume). Ultimately, it will ``take a village'' to develop the effective strategies required 
to explore theoretically, experimentally, and observationally the remaining open 
questions in this exciting field.

\section*{Acknowledgments}
I would like to thank my collaborators---too many to name here---for their
support and critical insights. This work was supported in part by a grant 
from the U.S. Department of Energy DE-FD05-92ER40750.

\bibliographystyle{spphys.bst}
\bibliography{SymmEnEJPA.bbl}

\begin{thebibliography}{100}
\providecommand{\url}[1]{{#1}}
\providecommand{\urlprefix}{URL }
\expandafter\ifx\csname urlstyle\endcsname\relax
  \providecommand{\doi}[1]{DOI \discretionary{}{}{}#1}\else
  \providecommand{\doi}{DOI \discretionary{}{}{}\begingroup
  \urlstyle{rm}\Url}\fi

\bibitem{Moller:1993ed}
P.\,M\"oller, J.R. Nix, W.D. Myers, and W.J. Swiatecki, 
Atom. Data Nucl. Data Tabl. \textbf{59},\,185\,(1995).

\bibitem{Moller:1997bz}
P.\,M\"oller, J.R. Nix, and K.L. Kratz, 
Atom. Data Nucl. Data Tabl. \textbf{66},\,131\,(1996).

\bibitem{Moller:2012}
P.\,M\"oller, W.D. Myers, H.\,Sagawa, and S.\,Yoshida, 
Phys. Rev. Lett. \textbf{108},\,052501\,(2012).

\bibitem{Duflo:1994}
J.\,Duflo, Nucl. Phys. \textbf{A576},\,29\,(1994).

\bibitem{Zuker:1994}
A.\,Zuker, Nucl. Phys. \textbf{A576},\,65\,(1994).

\bibitem{Duflo:1995}
J.\,Duflo and A.\,Zuker, Phys. Rev. C \textbf{52},\,R23\,(1995).

\bibitem{Harakeh:2001}
M.N. Harakeh and A.\,van\,der Woude, \emph{Giant Resonances-Fundamental
High-frequency Modes of Nuclear Excitation} (Clarendon, Oxford, 2001).

\bibitem{Paar:2007bk}
N.\,Paar, D.\,Vretenar E.\,Khan, and\,G.\,Col\`o, 
Rept. Prog. Phys.\,\textbf{70},\,691\,(2007).

\bibitem{Sagawa:2007pi}
H.\,Sagawa, S.\,Yoshida, X.R. Zhou, K.\,Yako, and H.\,Sakai, 
Phys. Rev. \textbf{C76},\,024301\,(2007).

\bibitem{Roca-Maza:2013yha}
X.\,Roca-Maza {\sl et\,al.,} Phys. Rev. \textbf{C87} 034301 (2013).

\bibitem{Piekarewicz:2008nh}
J.\,Piekarewicz and M.\,Centelles, 
Phys. Rev. \textbf{C79}, 054311 (2009).

\bibitem{Brown:2000}
B.A. Brown, Phys. Rev. Lett. \textbf{85}, 5296 (2000).

\bibitem{Furnstahl:2001un}
R.J. Furnstahl, Nucl. Phys. \textbf{A706}, 85 (2002).

\bibitem{Centelles:2008vu}
M.\,Centelles, X.\,Roca-Maza, X.\,Vi\~nas, and M.\,Warda, 
Phys. Rev. Lett. \textbf{102},\,122502\,(2009).

\bibitem{RocaMaza:2011pm}
X.\,Roca-Maza, M.\,Centelles, X.\,Vi\~nas, and M.\,Warda, Phys. 
Rev. Lett. \textbf{106}, 252501 (2011).

\bibitem{Hofstadter:1956qs}
R.\,Hofstadter, Rev. Mod. Phys. \textbf{28}, 214 (1956).

\bibitem{Donnelly:1989qs}
T.\,Donnelly, J.\,Dubach, and I.\,Sick, 
Nucl. Phys. \textbf{A503}, 589 (1989).

\bibitem{Angeli:2013}
I.\,Angeli and K.\, Marinova, At. Data Nucl. Data Tables \textbf{99}, 69  (2013).

\bibitem{Abrahamyan:2012gp}
S.\,Abrahamyan {\sl et\,al.,} Phys. Rev. Lett. \textbf{108}, 112502 (2012).

\bibitem{Horowitz:2012tj}
C.J.\,Horowitz {\sl et\,al.,}
Phys. Rev. \textbf{C85}, 032501 (2012).

\bibitem{Pollock:1992mv}
S.J. Pollock, E.N. Fortson, and L.\,Wilets, 
Phys. Rev. \textbf{C46}, 2587 (1992).

\bibitem{Sil:2005tg}
T.\,Sil, M.\,Centelles, X.\,Vi\~nas, and J.\,Piekarewicz, 
Phys. Rev. \textbf{C71}, 045502 (2005).

\bibitem{Guena:2005uj}
J.\,Guena, M.\,Lintz, and M.A. Bouchiat, 
Mod. Phys. Lett. \textbf{A20}, 375 (2005).

\bibitem{Behr:2008at}
J.\,Behr and G.\,Gwinner, 
J. Phys. \textbf{G36}, 033101 (2009).

\bibitem{Tsang:2004zz}
M.B. Tsang, {\sl et al.,} Phys. Rev. Lett. \textbf{92}, 062701 (2004).

\bibitem{Chen:2004si}
L.W. Chen, C.M. Ko, and B.A. Li, 
Phys. Rev. Lett. \textbf{94}, 032701 (2005).

\bibitem{Steiner:2005rd}
A.W. Steiner and B.A. Li, 
Phys. Rev. \textbf{C72}, 041601 (2005).

\bibitem{Shetty:2007zg}
D.V. Shetty, S.J. Yennello, and G.A. Souliotis, 
Phys. Rev. \textbf{C76}, 024606 (2007).

\bibitem{Tsang:2008fd}
M.B. Tsang  {\sl et al.,} Phys. Rev. Lett. \textbf{102}, 122701 (2009).

\bibitem{Horowitz:2000xj}
C.J. Horowitz and J.\,Piekarewicz, 
Phys. Rev. Lett. \textbf{86}, 5647 (2001).

\bibitem{Horowitz:2001ya}
C.J. Horowitz and J.\,Piekarewicz, 
Phys. Rev. \textbf{C64}, 062802 (2001).

\bibitem{Horowitz:2002mb}
C.J. Horowitz and J.\,Piekarewicz, 
Phys. Rev. \textbf{C66}, 055803 (2002).

\bibitem{Carriere:2002bx}
J.\,Carriere, C.J. Horowitz, and J.\,Piekarewicz, 
Astrophys. J. \textbf{593}, 463 (2003).

\bibitem{Steiner:2004fi}
A.W. Steiner, M.\,Prakash, J.M. Lattimer, and P.J. Ellis, 
Phys. Rept. \textbf{411}, 325 (2005).

\bibitem{Li:2005sr}
B.A. Li and A.W. Steiner, 
Phys. Lett. \textbf{B642}, 436 (2006).

\bibitem{Lattimer:2006xb}
J.M. Lattimer and M.\,Prakash, 
Phys. Rept. \textbf{442}, 109 (2007).

\bibitem{Fattoyev:2010tb}
F.J. Fattoyev and  J.\,Piekarewicz, 
Phys. Rev. \textbf{C82}, 025810 (2010).

\bibitem{Piekarewicz:2007dx}
J.\,Piekarewicz, Phys. Rev. \textbf{C76}, 064310 (2007).

\bibitem{Carbone:2010az}
A.\,Carbone {\sl et al.,}  
Phys. Rev. \textbf{C81}, 041301 (2010).

\bibitem{Hebeler:2010jx}
K.\,Hebeler, J.\,Lattimer, C.\,Pethick, and A.\,Schwenk, 
Phys. Rev. Lett. \textbf{105}, 161102 (2010).

\bibitem{Steiner:2011ft}
A.\,Steiner and  S.\,Gandolfi, 
Phys. Rev. Lett. \textbf{108}, 081102 (2012).

\bibitem{Hebeler:2013nza}
K.\, Hebeler, J.\,Lattimer, C.\, Pethick, and A.\, Schwenk,  
Astrophys. J. \textbf{773}, 11 (2013).
 
\bibitem{Tsang:2012se}
M.\, Tsang {\sl et al.,}  
Phys. Rev. \textbf{C86}, 015803 (2012).

\bibitem{Lattimer:2012nd}
J.M. Lattimer, Ann. Rev. Nucl. Part. Sci. \textbf{62}, 485 (2012).

\bibitem{Walecka:1974qa}
J.D. Walecka, Annals Phys. \textbf{83}, 491 (1974)

\bibitem{Serot:1979dc}
B.D. Serot, Phys. Lett. \textbf{B86}, 146 (1979).

\bibitem{Furnstahl:2000in}
R.J. Furnstahl and B.D. Serot, 
Comments Nucl. Part. Phys. \textbf{2}, A23 (2000).

\bibitem{Hohenberg:1964zz}
P.\,Hohenberg and W.\,Kohn, 
Phys. Rev. \textbf{136}, B864 (1964).

\bibitem{Kohn:1965}
W.\, Kohn and L.J. Sham, 
Phys. Rev. \textbf{140}, A1133 (1965).

\bibitem{Kohn:1999}
W.\,Kohn, Rev. Mod. Phys. \textbf{71}(5), 1253 (1999).

\bibitem{Furnstahl:1996wv}
R.J. Furnstahl, B.D. Serot, and H.B. Tang, 
Nucl. Phys. \textbf{A615}, 441 (1997).

\bibitem{Furnstahl:1996zm}
R.J. Furnstahl, B.D. Serot, and H.B. Tang, 
Nucl. Phys. \textbf{A618}, 446 (1997).

\bibitem{Rusnak:1997dj}
J.J. Rusnak and R.J. Furnstahl, 
Nucl. Phys. \textbf{A627}, 495 (1997).

\bibitem{Furnstahl:1997hq}
R.J. Furnstahl and J.C. Hackworth, Phys. Rev. \textbf{C56}, 2875 (1997).

\bibitem{Kortelainen:2010dt}
M.\,Kortelainen, R.J. Furnstahl, W.\,Nazarewicz, and M.\,Stoitsov, 
Phys. Rev. \textbf{C82}, 011304 (2010).

\bibitem{Mueller:1996pm}
H.\,Mueller and B.D. Serot, Nucl. Phys. \textbf{A606}, 508 (1996).

\bibitem{Horowitz:1981xw}
C.J. Horowitz and B.D. Serot, Nucl. Phys. \textbf{A368}, 503 (1981).

\bibitem{Serot:1984ey}
B.D. Serot and J.D. Walecka, Adv. Nucl. Phys. \textbf{16}, 1 (1986).

\bibitem{Boguta:1977xi}
J.\,Boguta and A.R. Bodmer, Nucl. Phys. \textbf{A292}, 413 (1977).

\bibitem{Youngblood:1999}
D.H. Youngblood, H.L. Clark, and Y.W. Lui, 
Phys. Rev. Lett. \textbf{82}(4), 691 (1999).

\bibitem{Lui:2004wm}
Y.W. Lui, D.H. Youngblood, Y.\,Tokimoto, H.L. Clark, and B.\,John, 
Phys. Rev. \textbf{C70}, 014307 (2004).

\bibitem{Uchida:2003}
M.\,Uchida {\sl et al.,} Phys. Lett. \textbf{B557}, 12  (2003).

\bibitem{Uchida:2004bs}
M.~Uchida {\sl et al.,}  Phys. Rev. \textbf{C69}, 051301 (2004).

\bibitem{Li:2007bp}
T.\,Li {\sl et al.,} Phys. Rev. Lett. \textbf{99}, 162503 (2007).

\bibitem{Li:2010kfa}
T.\,Li {\sl et al.,} Phys. Rev. \textbf{C81}, 034309 (2010).

\bibitem{Patel:2013}
U.\,Garg and D.\,Patel; {\sl private communication}.

\bibitem{Glendenning:2000}
N.K. Glendenning, \emph{Compact Stars} (Springer-Verlag New York, 2000).

\bibitem{Demorest:2010bx}
P.\,Demorest, T.\,Pennucci, S.\,Ransom, M.\,Roberts, and J.\,Hessels, 
Nature \textbf{467}, 1081 (2010).

\bibitem{Antoniadis:2013pzd}
J.\,Antoniadis {\sl et al.,}  Science \textbf{340}, 6131 (2013).

\bibitem{Todd:2003xs}
B.G. Todd and J.\,Piekarewicz, Phys. Rev. \textbf{C67}, 044317 (2003).

\bibitem{Fetter:1971}
A.L. Fetter and J.D. Walecka, 
\emph{Quantum Theory of Many Particle Systems} (McGraw-Hill, New York, 1971).

\bibitem{Dickhoff:2005}
W.H. Dickhoff and D.\,Van\,Neck, 
\emph{Many-body Theory Exposed} (World Scientific Publishing Co., 2005).

\bibitem{Dawson:1990wp}
J.F. Dawson and R.J. Furnstahl, Phys. Rev. \textbf{C42}, 2009 (1990).

\bibitem{Piekarewicz:2000nm}
J.\,Piekarewicz, Phys. Rev. \textbf{C62}, 051304 (2000).

\bibitem{Piekarewicz:2001nm}
J.\,Piekarewicz, Phys. Rev. \textbf{C64}, 024307 (2001).

\bibitem{Ma:2001hv}
Z.Y. Ma, A.\,Wandelt, N.\,Van\,Giai, D.\,Vretenar, P.\,Ring, and L.G. Cao,
Nucl. Phys. \textbf{A703}, 222 (2002).

\bibitem{Audi:2002rp}
G.\,Audi, A.H. Wapstra, and C.\,Thibault, 
Nucl. Phys. \textbf{A729}, 337 (2002).

\bibitem{Blaizot:1995}
J.P. Blaizot, J.F. Berger, J.\,Decharg\'e, and M.\,Girod, 
Nucl. Phys. \textbf{A591}, 435 (1995).

\bibitem{Piekarewicz:2003br}
J.\,Piekarewicz, Phys. Rev. \textbf{C69}, 041301 (2004).

\bibitem{Piekarewicz:2002jd}
J.\,Piekarewicz, Phys. Rev. \textbf{C66}, 034305 (2002).

\bibitem{Reinhard:2010wz}
P.G. Reinhard and  W.\,Nazarewicz, 
Phys. Rev. \textbf{C81}, 051303 (2010).

\bibitem{Piekarewicz:2010fa}
J.\,Piekarewicz, Phys.Rev. \textbf{C83}, 034319 (2011).

\bibitem{Piekarewicz:2012pp}
J.\,Piekarewicz {\sl et al.,} 
Phys. Rev. \textbf{C85}, 041302(R) (2012).

\bibitem{Roca-Maza:2013mla}
X.\,Roca-Maza {\sl et\,al.,} {\tt arXiv:1307.4806}\,[nucl-th]\,(2013).

\bibitem{Tamii:2011pv}
A.\,Tamii {\sl et\,al.,} Phys. Rev. Lett. \textbf{107}, 062502 (2011).

\bibitem{Poltoratska:2012nf}
I.\,Poltoratska  {\sl et\,al.,}
Phys. Rev. \textbf{C85} 041304 (2012). 

\bibitem{Satula:2005hy}
W.\,Satula, R.A. Wyss, and M.\,Rafalski, 
Phys. Rev. \textbf{C74}, 011301 (2006).

\bibitem{Lalazissis:1996rd}
G.A. Lalazissis, J.\,Konig, and P.\,Ring, 
Phys. Rev. \textbf{C55}, 540 (1997).

\bibitem{Lalazissis:1999}
G.A. Lalazissis, S.\,Raman, and P.\,Ring, 
At. Data Nucl. Data Tables \textbf{71}, 1 (1999).

\bibitem{Todd-Rutel:2005fa}
B.G. Todd-Rutel and J.\,Piekarewicz, 
Phys. Rev. Lett \textbf{95}, 122501 (2005).

\bibitem{Steiner:2010fz}
A.W. Steiner, J.M. Lattimer, and E.F. Brown, 
Astrophys. J. \textbf{722}, 33 (2010).

\bibitem{Fattoyev:2010mx}
F.J. Fattoyev, C.J. Horowitz, J.\,Piekarewicz, and G.\,Shen, 
Phys. Rev. \textbf{C82}, 055803 (2010).

\bibitem{Fattoyev:2013yaa}
F.J. Fattoyev and  J.\,Piekarewicz, 
{\tt arXiv:1306.6034}\,[nucl-th]\,(2013).

\bibitem{Chen:2013tca}
W.C.\,Chen, J.\,Piekarewicz, and M.\,Centelles,
Phys. Rev. C (in press); {\tt arXiv:1304.2421} [nucl-th]\,(2013).

\bibitem{Piekarewicz:2007us}
J.\,Piekarewicz, Phys. Rev. \textbf{C76}, 031301 (2007).

\bibitem{Sagawa:2007sp}
H.\,Sagawa, S.\,Yoshida, G.M. Zeng, J.Z. Gu, and X.Z. Zhang, 
Phys. Rev. \textbf{C76}, 034327 (2007).

\bibitem{Avdeenkov:2008bi}
A.\,Avdeenkov {\sl et al.,} Phys. Rev. \textbf{C79}, 034309 (2009).

\bibitem{Piekarewicz:2009gb}
J.\,Piekarewicz, J. Phys. \textbf{G37}, 064038 (2010).

\bibitem{Cao:2012dt}
L.G. Cao, H.\,Sagawa, G.\,Col\`o,  
Phys. Rev.\textbf{ C86} 054313   (2012). 

\bibitem{Patel:2012zd}
D.\,Patel  {\sl et\,al.,} Phys. Lett. \textbf{B718}, 447 (2012).

\bibitem{Li:2008hx}
J.\,Li, G.\,Col\`o, and J.\,Meng, 
Phys. Rev. \textbf{C78}, 064304 (2008).

\bibitem{Khan:2009xq}
E.\,Khan, Phys. Rev. \textbf{C80}, 011307 (2009).

\bibitem{Khan:2009ih}
E.\,Khan, Phys. Rev. \textbf{C80}, 057302 (2009).

\bibitem{Khan:2010mv}
E. Khan, J. Margueron, G. Col\`o, K. Hagino, and H.\,Sagawa, 
Phys. Rev. \textbf{C82}, 024322 (2010).

\bibitem{Vesely:2012dw}
P. Vesely, J. Toivanen, B.G. Carlsson, J. Dobaczewski, N. Michel, and A. Pastore, 
Phys. Rev. \textbf{C86}, 024303 (2012). 

\bibitem{RocaMaza:2012sj}
X.\,Roca-Maza, G. Col\`o, and H.\,Sagawa, 
Phys. Rev. \textbf{C86}, 031306 (2012).

\bibitem{Suzuki:1990}
Y. Suzuki, K. Ikeda, and H. Sato, 
Prog. Theor. Phys. \textbf{83}, 180 (1990).

\bibitem{VanIsacker:1992}
P.\,Van\,Isacker, D.D. Nagarajan, and M.A. Warner, 
Phys. Rev. \textbf{C45}, R13 (1992).

\bibitem{Hamamoto:1996}
I. Hamamoto, H. Sagawa, and X.Z. Zhang, 
Phys. Rev. \textbf{C53}, 765 (1996).

\bibitem{Hamamoto:1998}
I. Hamamoto, H. Sagawa, and X.Z. Zhang, 
Phys. Rev. \textbf{C57}, R1064 (1998).

\bibitem{Vretenar:2000yy}
D. Vretenar, N. Paar, P. Ring, and G.A. Lalazissis, 
Phys. Rev. \textbf{C63}, 047301 (2001).

\bibitem{Vretenar:2001hs}
D. Vretenar, N. Paar, P. Ring, G.A. Lalazissis, 
Nucl. Phys. \textbf{A692}, 496 (2001).

\bibitem{Paar:2002gz}
N. Paar, P. Ring, T. Niksic, and D. Vretenar, 
Phys. Rev. \textbf{C67}, 034312 (2003).

\bibitem{Tsoneva:2003gv}
N. Tsoneva, H. Lenske, and C. Stoyanov, 
Phys. Lett. \textbf{B586}, 213 (2004).

\bibitem{Sarchi:2004}
D. Sarchi, P.F. Bortignon, and G. Col\`o, 
Phys. Lett. \textbf{B601}, 27 (2004).

\bibitem{Paar:2004gr}
N. Paar, T. Niksic, D. Vretenar, and P. Ring, 
Phys. Lett. \textbf{B606}, 288 (2005).

\bibitem{Piekarewicz:2006ip}
J.\,Piekarewicz, Phys. Rev. \textbf{C73}, 044325 (2006).

\bibitem{Tsoneva:2007fk}
N. Tsoneva and H. Lenske, 
Phys. Rev. \textbf{C77}, 024321 (2008).

\bibitem{Klimkiewicz:2007zz}
A. Klimkiewicz {\sl et al.,} Phys. Rev. \textbf{C76}, 051603 (2007).

\bibitem{Adrich:2005}
P. Adrich {\sl et al.,} Phys. Rev. Lett \textbf{95}, 132501 (2005).

\bibitem{Wieland:2009}
O. Wieland {\sl et al.,} Phys. Rev. Lett. \textbf{102}(9), 092502 (2009).

\bibitem{Rossi:2012ew}
D. Rossi {\sl et al.,} J. Phys. Conf. Ser. \textbf{420}, 012072 (2013).

\bibitem{Savran:2013bha}
D. Savran, T. Aumann, and A.~Zilges, 
Prog. Part. Nucl. Phys. \textbf{70}, 210 (2013).

\bibitem{UNEDF}
Building a universal nuclear energy density functional.
\newblock \urlprefix\url{http://unedf.org}.
\newblock (UNEDF Collaboration).

\bibitem{Kortelainen:2010hv}
M. Kortelainen {\sl et al.,}  Phys. Rev. \textbf{C82}, 024313 (2010).

\bibitem{Fattoyev:2011ns}
F.J Fattoyev and J.\,Piekarewicz, Phys. Rev. \textbf{C84}, 064302 (2011).

\bibitem{Fattoyev:2012rm}
F.J. Fattoyev and  J.\,Piekarewicz, Phys. Rev. \textbf{C88}, 015802 (2012).

\bibitem{Erler:2012qd}
J. Erler, C.J. Horowitz, W. Nazarewicz, M. Rafalski, and P.G. Reinhard,   
{\tt arXiv:1211.6292} [nucl-th]\,(2012).

\end{thebibliography}

\end{document}